\documentclass[aps,prd,superscriptaddress,showpacs,nofootinbib]{revtex4}
\usepackage{graphicx,epsfig,wrapfig}
\newcommand{\be}{\begin{equation}}
\newcommand{\ee}{\end{equation}}
\newcommand{\ba}{\begin{eqnarray}}
\newcommand{\ea}{\end{eqnarray}}
\newcommand{\di}{\!{\rm d}}
\newcommand{\la}{\langle}
\newcommand{\ra}{\rangle}
\newcommand{\fslash}[1] {{\not\! #1\,}}
\newcommand{\bDelta}{ {\bf\Delta}}
\newcommand{\kringel}[1]{ \;\stackrel{\circ}{#1}}
\newcommand{\doublesum}[2]
    {\!\sum_{{{\scriptscriptstyle {#1}}\atop{\scriptscriptstyle {#2}}}}}

\begin{document}
\newcommand*{\Bochum}{Institut f\"ur Theoretische Physik II,
    Ruhr-Universit\"at Bochum, D-44789 Bochum, Germany}\affiliation{\Bochum}
\newcommand*{\Coimbra}{Centro de F\'isica Computacional (CFC),
    Departamento de F\'isica, Universidade de Coimbra,
    P-3004-516  Coimbra, Portugal}\affiliation{\Coimbra}
\newcommand*{\Porto}{Faculdade de Engenharia da Universidade do Porto,
    P-4200-465 Porto, Portugal}\affiliation{\Porto}
\title{
    The pion mass dependence of the nucleon form-factors \\
    of the energy momentum tensor in the chiral quark-soliton model}
\author{K.~Goeke}\affiliation{\Bochum}
\author{J.~Grabis}\affiliation{\Bochum}
\author{J.~Ossmann}\affiliation{\Bochum}
\author{P.~Schweitzer}\affiliation{\Bochum}
\author{A.~Silva}\affiliation{\Coimbra}\affiliation{\Porto}
\author{D.~Urbano}\affiliation{\Porto}\affiliation{\Coimbra}
\date{December, 2006}
\begin{abstract}
The nucleon form factors of the energy-momentum tensor are studied in
the large-$N_c$ limit in the framework of the chiral quark-soliton model
for model parameters that simulate physical situations in which pions
are heavy. This allows for a direct comparison to lattice QCD results.
\end{abstract}
\pacs{
  11.15.Pg, 
  12.39.Fe, 
  12.38.Lg, 
  12.38.Gc} 



\maketitle

  \vspace{-7.4cm}
  \begin{flushright}
  preprint RUB-TP2-08-2006
  \end{flushright}
  \vspace{5.3cm}

\section{Introduction}
\label{Sec-1:introduction}

The nucleon energy momentum tensor (EMT) form factors \cite{Pagels}
contain valuable information on the nucleon structure. They carry information
on, e.g., how the quark and gluon degrees of freedom share the total
momentum, and angular momentum of the nucleon \cite{Ji:1996ek}, or on the
distribution of strong forces inside the nucleon \cite{Polyakov:2002yz}.
The first is known from deeply inelastic lepton nucleon scattering experiments.
The latter two can be deduced from generalized parton distribution functions
(GPDs) \cite{Muller:1998fv,Ji:1996nm,Collins:1996fb,Radyushkin:1997ki}
accessible in hard exclusive reactions
\cite{Saull:1999kt,Adloff:2001cn,Airapetian:2001yk,Stepanyan:2001sm,Ellinghaus:2002bq,Chekanov:2003ya}, see
\cite{Ji:1998pc,Radyushkin:2000uy,Goeke:2001tz,Diehl:2003ny,Belitsky:2005qn}
for reviews.

Nucleon EMT form factors were studied in lattice QCD calculations
\cite{Mathur:1999uf,Gadiyak:2001fe,Hagler:2003jd,Gockeler:2003jf,Negele:2004iu}.
In principle, lattice QCD provides a rigorous and model-independent
approach to compute the nucleon EMT form factors.
In practice, however, present day technics and computing power allow
to simulate on lattices ``worlds'' with pions of typically
$m_\pi\gtrsim 400\,{\rm MeV}$.
The situation is extected to improve in the future, see
\cite{Schroers:2007qf} for status reports on selected
topics.

For the time being, however, it is necessary to use chiral extrapolation
in order to relate lattice results to the real world situation.
Chiral perturbation theory ($\chi$PT) provides a model-independent tool for that,
and the chiral behaviour of the nucleon EMT form factors was studied in
\cite{Chen:2001pv,Belitsky:2002jp,Diehl:2006ya}.
Experience with chiral extrapolations of other nucleon properties
indicates that $\chi$PT is applicable up to the lowest presently
available lattice values of $m_\pi$ \cite{Bernard:2003rp,Procura:2003ig,AliKhan:2003cu}
although the issue is not yet settled \cite{Beane:2004ks}.

In this situation it is worth looking on what one can learn about
the chiral behaviour of nucleon properties from other effective approaches,
e.g.\  the ``finite range regulator'' (FRR) approach. There chiral loops are
regulated by suitably chosen vertex form factors in order to phenomenologically
simulate the effects of the pion cloud which has a finite range due to $m_\pi\neq 0$
\cite{Leinweber:1998ej,Thomas:2005rz}. However, model calculations
\cite{Matevosyan:2005fz,Schweitzer:2003sb,Goeke:2005fs}
are equally of interest in this context.

A phenomenologically successful and theoretically consistent model for the description
of nucleon properties {\sl at the physical point} and {\sl in the chiral limit},
is the chiral quark soliton model (CQSM) \cite{Diakonov:yh,Diakonov:1987ty}.
This model describes the nucleon as a soliton of a static background pion field
in the limit of a large number of colours $N_c$, and hence provides a particular
realization of the general large-$N_c$ picture of the nucleon \cite{Witten:1979kh}.
The CQSM describes numerous nucleonic properties without adjustable parameters
-- including among others form factors
\cite{Christov:1995hr,Christov:1995vm,Silva:2005qm},
usual parton distribution functions
\cite{Diakonov:1996sr,Diakonov:1997vc,Diakonov:1998ze,Wakamatsu:1998rx,Goeke:2000wv}
and GPDs \cite{Petrov:1998kf,Penttinen:1999th,Schweitzer:2002nm,Schweitzer:2003ms,Ossmann:2004bp,Wakamatsu:2005vk}
--- within an accuracy of $(10-30)\%$ as far as those quantities are known.

That it is possible to extend the CQSM to the description of the nucleon
at {\sl large pion masses} was shown in \cite{Goeke:2005fs},
where the model was demonstrated to provide a good description of lattice data on the
$m_\pi$-dependence of the nucleon mass $M_N$ up to $m_\pi\lesssim 1.5\,{\rm GeV}$.
An important prerequisite for that is that the CQSM formally contains the
correct heavy quark limit result for the nucleon mass $M_N$ \cite{Goeke:2005fs}.

In this work we present a study of the $m_\pi$-dependence of the nucleon EMT form
factors in the CQSM up to pion masses as large as $1\,{\rm GeV}$.
The present study extends the study of Ref.~\cite{accompanying-paper-I}
where the nucleon EMT form factors were studied at the physical point
and in the chiral limit, and its purpose is threefold.

First, we provide an important supplement for the study in Ref.~\cite{Goeke:2005fs}.
There soliton solutions were obtained numerically for model parameters corresponding
to pion masses in the range $0\le m_\pi\le 1.5\,{\rm GeV}$. Here we provide a
cross check demonstrating that the numerical solutions found
in \cite{Goeke:2005fs} correspond,  in fact, to stable solitons.

Second, with the results obtained for large $m_\pi$ we are in the position
to confront the model predictions for the nucleon EMT form factors directly to
lattice QCD results
\cite{Mathur:1999uf,Gadiyak:2001fe,Hagler:2003jd,Gockeler:2003jf,Negele:2004iu}.
In view of the early stage of art of the experimental situation of hard exclusive
reactions, such a comparison provides the only presently available test for
our results.

Third, though the model --- as discussed in detail in \cite{Goeke:2005fs} ---
cannot be used as a quantitative guideline for the chiral extrapolation, our study
still allows to gain several interesting qualitative insights with this respect.

The note is organized as follows.
In Sec.~\ref{Sec-2:EMT-in-general} we introduce the nucleon EMT form factors and
discuss their properties.
In Sec.~\ref{Sec-3:model} we briefly review how the nucleon EMT form factors
are described in the CQSM.
In Secs.~\ref{Sec-4:densities} and \ref{Sec-5:form-factors} we describe respectively
the model results for the densities associated with the form factors and the
form factors themselves.
In Sec.~\ref{Sec-6:lattice} we compare the model results with lattice QCD data,
and in Sec.~\ref{Sec-7:lattice-lessons} we discuss which qualitative observations
from our study could be of interest in the context of the chiral extrapolation
of lattice data. Sec.~\ref{Sec-7:conclusions} contains the conclusions.
A remark on different notations for the EMT form factors is posed
in App.~\ref{App:Alternative-definition}.

\section{Form factors of the energy-momentum tensor}
\label{Sec-2:EMT-in-general}

The nucleon matrix element of the symmetric EMT
of QCD is characterized by three scalar form factors \cite{Pagels}.
The quark and gluon parts, $\hat T_{\mu\nu}^Q$ and $\hat T_{\mu\nu}^G$,
of the EMT are separately gauge-invariant and can be parameterized as
\cite{Ji:1996ek,Polyakov:2002yz},
see App.~\ref{App:Alternative-definition} for an alternative notation,
\ba
    \la p^\prime| \hat T_{\mu\nu}^{Q,G}(0) |p\rangle
    &=& \bar u(p^\prime)\biggl[M_2^{Q,G}(t)\,\frac{P_\mu P_\nu}{M_N}+
    J^{Q,G}(t)\ \frac{i(P_{\mu}\sigma_{\nu\rho}+P_{\nu}\sigma_{\mu\rho})
    \Delta^\rho}{2M_N}
    \nonumber \\
    &+& d^{Q,G}_1(t)\,
    \frac{\Delta_\mu\Delta_\nu-g_{\mu\nu}\Delta^2}{5M_N}
    \pm \bar c(t)g_{\mu\nu} \biggr]u(p)\, .
    \label{Eq:ff-of-EMT} \ea
Here the nucleon states and spinors are normalized by
$\la p^\prime|p\ra = 2p^0(2\pi)^3\delta^{(3)}({\bf p}^\prime-{\bf p})$
and $\bar u(p) u(p)=2 M_N$, and spin indices are suppressed for brevity.
The kinematical variables are defined as $P=(p+p')/2$, $\Delta=(p'-p)$,
$t=\Delta^2$.
The form factor $\bar c(t)$ accounts for non-conservation of the separate
quark and gluon parts of the EMT, and enters the quark and gluon parts
with opposite signs such that the total (quark+gluon) EMT is conserved.

The nucleon form factors of the EMT are related to the second Mellin
moments of the unpolarized GPDs $H^f(x,\xi,t)$ and $E^f(x,\xi,t)$ as
(we use the notation of Ref.~\cite{Goeke:2001tz})
\be
    \int\limits_{-1}^1 \!dx\; x\, \sum\limits_f H^f(x,\xi,t)
        = M_2^Q(t) +\frac 45\, d_1^Q(t)\, \xi^2 \;,\;\;\;\;
        \int\limits_{-1}^1 \!dx\; x\, \sum\limits_f (H^f+E^f)(x,\xi,t)
        = 2J^Q(t)\, .
        \label{Eq:EMT-and-GPD-H-E} \ee
where $\xi$ denotes the so-called skewedness parameter \cite{Ji:1996ek}.
The sum rules in Eqs.~(\ref{Eq:EMT-and-GPD-H-E}) are special cases of
the so-called polynomiality property of GPDs \cite{Ji:1998pc}.
The second sum rule in  (\ref{Eq:EMT-and-GPD-H-E}) provides the possibility to access
$J^Q(0)$, i.e.\  the total (spin+orbital angular momentum) contribution of quarks to
the nucleon spin, through the extraction of GPDs from hard exclusive processes and
extrapolation to the unphysical point $t=0$.
The sensitivity of different observables to the total quark angular momenta
was investigated in model studies \cite{Goeke:2001tz,Ellinghaus:2005uc}.
For gluons there are analog definitions and expressions.
Suffice to remark that the full GPDs contain far more
information \cite{Ref-further-information-in-GPDs}.

The form factors of the EMT in Eq.~(\ref{Eq:ff-of-EMT}) can be interpreted
\cite{Polyakov:2002yz} in analogy to the electromagnetic form factors
\cite{Sachs} in the Breit frame characterized by $\Delta^0=0$.
In this frame one can define the static energy-momentum tensor for quarks
\be\label{Def:static-EMT}
    T_{\mu\nu}^Q({\bf r},{\bf s}) =
    \frac{1}{2E}\int\frac{\di^3\bDelta}{(2\pi)^3}\;\exp(i\bDelta{\bf r})\;
    \la p^\prime,S^\prime|\hat{T}_{\mu\nu}^Q(0)|p,S\ra \, ,
\ee
and analogously for gluons. The initial and final polarization vectors of the
nucleon $S$ and $S^\prime$ are defined such that in the respective rest-frame
they are equal to $(0,{\bf s})$ with the unit vector ${\bf s}$ denoting
the quantization axis for the spin.

The components of $T_{0 k}^Q({\bf r},{\bf s})$ and
$\varepsilon^{i j k} r_j T_{0k}^Q({\bf r},{\bf s})$ correspond respectively
to the distribution of quark momentum and quark angular momentum inside the
nucleon.
The components of $(T_{ik}^Q-\frac 13\delta_{ik}T^Q_{ll})({\bf r},{\bf s})$
characterize the spatial distribution of ``shear forces" experienced by
quarks inside the nucleon. The respective form factors are related to
$T_{\mu\nu}^Q({\bf r},{\bf s})$ by
\ba
    J^Q(t)+\frac{2t}{3}\, {J^Q}^\prime(t)
    &=& \int\di^3{\bf r}\, e^{-i{\bf r}\bDelta}\,
    \varepsilon^{ijk}\,s_i\,r_j\,T_{0k}^Q({\bf r},{\bf s})\, ,
    \label{Eq:ff-J}\\
    d_1^Q(t)+\frac{4t}{3}\, {d_1^Q}^\prime(t)
    +\frac{4t^2}{15}\, {d_1^Q}^{\prime\prime}(t)
    &=& -\frac{M_N}{2}\, \int\di^3{\bf r}\,e^{-i{\bf r}\bDelta}\,
    T_{ij}^Q({\bf r})\,\left(r^i r^j-\frac{{\bf r}^2}3\,\delta^{ij}\right)\,
    , \label{Eq:ff-d1}\\
    M_2(t)-\frac{t}{4M_N^2}\left(M_2(t)-2 J(t)+\frac 45\, d_1(t) \right)
    &=&\frac{1}{M_N}\,\int\di^3{\bf r}\, e^{-i{\bf r}\bDelta}
    \, T_{00}({\bf r},{\bf s})\,,
    \label{Eq:ff-M2}
\ea
where the prime denotes derivative with respect to the Mandelstam variable $t$.
Note that for a spin-1/2 particle only the $T^{0\mu}$-components are sensitive
to the polarization vector. Note also that Eq.~(\ref{Eq:ff-M2}) holds for the
sum $T_{00}\equiv T_{00}^Q+T_{00}^G$ with $M_2(t)\equiv M_2^Q(t)+M_2^G(t)$ and
$J(t)$ and $d_1(t)$ defined analogously, but not for the separate
quark and gluon contributions -- since otherwise the form factor
$\bar c(t)$ would not cancel out.

The form factor $M_2(t)$ at $t=0$ is connected to the fractions of the
nucleon momentum carried respectively by quarks and gluons. More precisely
\be\label{Eq:M2-at-t=0}
    M_2^Q(0) = \int\limits_0^1\!\di x\;
               \sum\limits_q x(f_1^q+f_1^{\bar q})(x)\;,\;\;\;
    M_2^G(0) = \int\limits_0^1\!\di x\;xf_1^g(x)\;,
\ee
where $f_1^a(x)=H^a(x,0,0)$ are the unpolarized parton distributions
accessible in inclusive deeply inelastic scattering.

The form factors $M_2^{Q,G}(t)$, $J^{Q,G}(t)$ and $d_1^{Q,G}(t)$ are
renormalization scale dependent (the indication of the renormalization
scale $\mu$ is suppressed for brevity). Their quark+gluon sums, however, are
scale independent form factors, which at $t=0$ satisfy the constraints,
\ba
    M_2(0)&=&
    \frac{1}{M_N}\,\int\di^3{\bf r}\;T_{00}({\bf r},{\bf s})=1
    \;,\nonumber\\
    J(0)&=&
    \int\di^3{\bf r}\;\varepsilon^{ijk}\,s_i\,r_j\,
    T_{0k}({\bf r},{\bf s})=\frac12\, , \nonumber\\
    d_1(0)&=&
    -\frac{M_N}{2}\, \int\di^3{\bf r}\;T_{ij}({\bf r})\,
    \left(r^i r^j-\frac{{\bf r}^2}3\,\delta^{ij}\right)\equiv d_1\,,
    \label{Eq:M2-J-d1}
\ea
which mean that in the rest frame the total energy of the nucleon is equal
to its mass, and that its spin is 1/2. The value of $d_1$
is not known a priori and must be determined experimentally. However,
being a conserved quantity it is to be considered on the same footing as
other basic nucleon properties like mass, anomalous magnetic moment, etc.
Remarkably, $d_1$ determines the behaviour of the $D$-term
\cite{Polyakov:1999gs} (and thus the unpolarized GPDs) in the asymptotic
limit of renormalization scale $\mu\to\infty$ \cite{Goeke:2001tz}.

The form factor $d_1(t)$ is connected to the distribution of pressure and
shear forces experienced by the partons in the nucleon \cite{Polyakov:2002yz}
which becomes apparent by recalling that $T_{ij}({\bf r})$ is the static
stress tensor which (for spin 0 and 1/2 particles) can be decomposed as
\be\label{Eq:T_ij-pressure-and-shear}
    T_{ij}({\bf r})
    = s(r)\left(\frac{r_ir_j}{r^2}-\frac 13\,\delta_{ij}\right)
        + p(r)\,\delta_{ij}\, . \ee
Hereby $p(r)$ describes the radial distribution of the ``pressure" inside the
hadron, while $s(r)$ is related to the distribution of the ``shear forces''
\cite{Polyakov:2002yz}. Both are related due to the conservation
of the EMT by the differential equation
\be\label{Eq:p(r)+s(r)}
    \frac23\;\frac{\partial s(r)}{\partial r\;}+
    \frac{2s(r)}{r} + \frac{\partial p(r)}{\partial r\;} = 0\;.
\ee
 Another important consequence of the conservation of the EMT
is the so-called stability condition
\be\label{Eq:stability}
    \int\limits_0^\infty \!\di r\;r^2p(r)=0 \;.
\ee

Let us review briefly what is known about $d_1$
--- which in terms of the pressure or shear forces is given by
\be\label{Eq:d1-from-s(r)-and-p(r)}
        d_1  =  -\,\frac{1}{3}\;M_N \int\di^3{\bf r}\;r^2\, s(r)
         =     \frac{5}{4}\;M_N \int\di^3{\bf r}\;r^2\, p(r)\;.
\ee
For the pion it can be calculated exactly using soft pion theorems
with the result $d_{1,\pi}^Q=-M_{2,\pi}^Q$ \cite{Polyakov:1999gs},
see also \cite{Teryaev:2001qm}.
Also for the nucleon $d_1^Q < 0$ was found in calculations in CQSM
\cite{accompanying-paper-I,Petrov:1998kf,Kivel:2000fg}.
For the nucleon the large-$N_c$ limit predicts the flavour-dependence
$|d_1^u+d_1^d| \gg |d_1^u-d_1^d|$ \cite{Goeke:2001tz}. Lattice calculations
\cite{Mathur:1999uf,Gadiyak:2001fe,Hagler:2003jd,Gockeler:2003jf,Negele:2004iu}
confirm this
flavour dependence and yield a negative $d_1^Q$, see Sec.~\ref{Sec-6:lattice}.
In a simple ``liquid drop'' model $d_1$ can be related to the surface tension
of the ``liquid'' and comes out negative \cite{Polyakov:2002yz}.
Such a model is in particular applicable to nuclei and predictions from this picture
\cite{Polyakov:2002yz} were confirmed in calculations assuming realistic nuclear
models \cite{Guzey:2005ba}.
In Ref.~\cite{accompanying-paper-I} it was conjectured on the basis of plausible
physical arguments that the negative sign of $d_1$ is dictated by stability criteria.
This conclusion, however, remains to be proven for the general case.

\newpage
\section{Nucleon EMT form factors in the CQSM}
\label{Sec-3:model}

The effective theory underlying the CQSM was derived from the instanton model
of the QCD vacuum \cite{Diakonov:1983hh,Diakonov:1985eg,Diakonov:1995qy}
which assumes that the basic properties of the QCD vacuum are dominated
by a strongly interacting but dilute instanton medium, see the reviews
\cite{Diakonov:2000pa}. In this medium light quarks acquire a dynamical
(``constituent'') quark mass due to interactions with instantons.
At low momenta below a scale set by $\rho_{\rm av}^{-1}\approx 600\,{\rm MeV}$,
where $\rho_{\rm av}$ denotes the average instanton size, the dynamics of the
effective quark degrees of freedom is given by the partition function
\cite{Diakonov:1984tw,Dhar:1985gh}
\be\label{eff-theory}
    Z_{\rm eff} = \int\!\!{\cal D}\psi\,{\cal D}\bar{\psi}\,
    {\cal D}U\,\exp\biggl(iS_{\rm eff}(\bar\psi,\psi,U)\biggr)\;,\;\;\;
    S_{\rm eff}(\bar\psi,\psi,U)=\int\di^4x\;\bar{\psi}\,
    (i\fslash{\partial}-M\,U^{\gamma_5}-m)\psi\,.\ee
Here we restrict ourselves to two flavours and neglect isospin breaking effects
with $m=m_u=m_d$ denoting the current quark mass, while $U=\exp(i\tau^a\pi^a)$
denotes the chiral pion field with $U^{\gamma_5} = \exp(i\gamma_5\tau^a\pi^a)$.
The dynamical mass is strictly speaking momentum dependent, i.e.\ $M=M(p)$.
However, in practical calculations it is convenient to work with a constant
$M=M(0)=350\,{\rm MeV}$ following from the instanton vacuum \cite{Diakonov:2000pa},
and to regularize the effective theory by means of an explicit regularization
with a cutoff of ${\cal O}(\rho_{\rm av}^{-1})$ whose precise value is fixed
to reproduce the physical value of the pion decay constant $f_\pi=93\,{\rm MeV}$.
In this work we will use the proper-time regularization.

The quark degrees of freedom of the effective theory (\ref{eff-theory}) correspond
to QCD quark degrees of freedom up to corrections which are small in the instanton
packing fraction $\rho_{\rm av}/R_{\rm av}\sim\frac13$, where $R_{\rm av}$
denotes the average separation of instantons. The same parameter suppresses
the contribution of gluon degrees of freedom \cite{Diakonov:1995qy}.

The CQSM is an application of the effective theory (\ref{eff-theory})
to the description of baryons \cite{Diakonov:yh,Diakonov:1987ty}.
While the Gaussian path integral over fermion fields in (\ref{eff-theory})
can be solved exactly, the path integral over pion field configurations can be
solved only in the large-$N_c$ limit by means of the saddle-point approximation
(in the Euclidean formulation of the theory).
In the leading order of the large-$N_c$ limit the pion field is static,
and one can determine the spectrum of the one-particle Hamiltonian of the
effective theory (\ref{eff-theory})
\be\label{Hamiltonian}
    \hat{H}|n\ra=E_n |n\ra \;,\;\;
    \hat{H}=-i\gamma^0\gamma^k\partial_k+\gamma^0MU^{\gamma_5}+\gamma^0m
    \;. \ee
The spectrum of (\ref{Hamiltonian}) consists of an upper and a lower Dirac continuum,
distorted by the pion field as compared to continua of the free Dirac-Hamiltonian
$\hat{H}_0$ (given by $\hat{H}$ with $U^{\gamma_5}$ replaced by $1$),
and of a discrete bound state level of energy $E_{\rm lev}$,
if the pion field is strong enough.
By occupying the discrete level and the lower continuum states each by $N_c$ quarks
in an anti-symmetric colour state, one obtains a state with unity baryon number.
The soliton energy
\be\label{Eq:soliton-energy}
    E_{\rm sol}[U] = N_c \biggl[E_{\rm lev}+
    \sum\limits_{E_n<0}(E_n-E_{n_0})\biggr]_{\rm reg} \;.
    \ee
is a functional of the pion field. It is logarithmically divergent, see e.g.\
\cite{Christov:1995vm} for explicit expressions in the proper-time regularization.
By minimizing $E_{\rm sol}[U]$ one obtains the self-consistent solitonic
pion field $U_c$.  This procedure is performed for symmetry reasons in
the so-called hedgehog ansatz $\pi^a({\bf x})=e^a_r\;P_c(r)$
with the radial (soliton profile) function $P_c(r)$ and $r=|{\bf x}|$,
${\bf e}_r = {\bf x}/r$.
The nucleon mass $M_N$ is given by $E_{\rm sol}[U_c]$.

In the large-$N_c$ limit the path integral over $U$ in Eq.~(\ref{eff-theory}) is
solved by evaluating the expression at $U_c$ and integrating over translational
and rotational zero modes of the soliton solution.
In order to include corrections in the $1/N_c$-expansion one considers time dependent
pion field fluctuations around the solitonic solution. In practice hereby one
restricts oneself to time dependent rotations of the soliton field in spin- and
flavour-space which are slow because the corresponding soliton moment of inertia
\be\label{Eq:mom-inertia}
    I=\frac{N_c}{6}\doublesum{m,\rm non}{n,\rm occ}
    \frac{\la n|\tau^a|m\ra\,\la m|\tau^a|n\ra}{E_m-E_n}\biggl|_{\rm reg}
\ee
is large, $I={\cal O}(N_c)$.
It is logarithmically divergent and has to be regularized.
In (\ref{Eq:mom-inertia}) one has to sum over occupied (``occ'') states $n$
which satisfy $E_n\le E_{\rm lev}$, and over non-occupied (``non'') states
$m$ which satisfy $E_m > E_{\rm lev}$.

The model expressions for the nucleon EMT form factors in the effective
theory (\ref{eff-theory}) were derived explicitly in \cite{accompanying-paper-I}.
The gluon part of the EMT is zero in the effective theory (\ref{eff-theory}),
because there are no explicit gluon degrees of freedom.
(So we omit the index $Q$ when discussing the model results in this and
in Secs.~\ref{Sec-4:densities},~\ref{Sec-5:form-factors} but restore it later.)

Consequently, in the model the quark part of the EMT is conserved by itself,
and the form-factor $\bar c(t)$ in Eq.~(\ref{Eq:ff-of-EMT}) vanishes
\cite{accompanying-paper-I}.
The model expressions for the other form factors read
\newpage
\ba
        M_2(t)-\frac{t}{5M_N^2}d_1(t)
        &=& \frac{1}{M_N}\int\di^3{\bf r}\;\rho_E(r) \;j_0(r\sqrt{-t})
        \label{Eq:M2-d1-model-comp}\\
        d_1(t)
        &=& \frac{15 M_N}{2}\int\di^3{\bf r}\;p(r) \;\frac{j_0(r\sqrt{-t})}{t}
        \label{Eq:d1-model-comp}\\
        J(t)
        &=& 3
    \int\di^3{\bf r}\;\rho_J(r)\;\frac{j_1(r\sqrt{-t})}{r\sqrt{-t}}\;,
    \label{Eq:J-model-comp}
\ea
with the Bessel functions $j_0(z) = \frac{\sin z}{z}$ and $j_1(z)=-j_0^\prime(z)$.
The Fourier transforms of the form factors,  which are radial functions and to
which we refer as ``densities'' in the following, are defined as
\ba
    \rho_E(r)
    &=&
    N_c\sum\limits_{n,\rm occ}E_n\;
        \phi_n^\ast({\bf r})\phi_n({\bf r})\biggl|_{\rm reg}
    \label{Eq:density-energy}\\
    p(r)
    &=&
    \frac{N_c}{3}\sum_{n,\rm occ}\phi_n^\ast({\bf r})\,(\gamma^0
    \mbox{\boldmath$\gamma$}\hat{\bf p})\,\phi_n({\bf r})\biggl|_{\rm reg}
    \label{Eq:density-pressure}\\
    \rho_J(r)
    &=&
    - \frac{N_c}{24I} \doublesum{n,\rm occ}{j,\rm non}\epsilon^{abc}r^a
    \phi_j^\ast({\bf r})\biggl(2\hat{p}^b+(E_n+E_j)\gamma^0\gamma^b\biggr)
    \phi_n({\bf r})\;\frac{\langle n|\tau^c|j\rangle}{E_j-E_n}\,
    \biggl|_{\rm reg}\;.
    \label{Eq:density-spin}
\ea
The expressions in
(\ref{Eq:density-energy},~\ref{Eq:density-pressure},~\ref{Eq:density-spin})
are logarithmically UV-divergent. Here we use the proper-time method to regularize
them, see \cite{accompanying-paper-I} for explicit expressions in this regularization.
In Ref.~\cite{accompanying-paper-I} analytical proofs were given that
\begin{itemize}
\item   the stability condition (\ref{Eq:stability}) is satisfied in the model,
\item   the form factors satisfy the constraints at $t=0$ in
    Eq.~(\ref{Eq:M2-J-d1}), and
\item   the same expressions for EMT form factors follow in the model from
    unpolarized GPDs via the sum rules (\ref{Eq:EMT-and-GPD-H-E}).
\end{itemize}
Notice that the latter is a special case of the ``polynomiality
property'' of GPDs \cite{Ji:1998pc} satisfied in the CQSM
\cite{Schweitzer:2002nm,Schweitzer:2003ms}.
The field-theoretic character of the model is a crucial prerequisite
which allows to formulate and analytically prove such and other
\cite{Diakonov:1996sr} general QCD requirements.
This in turn provides important cross checks for the theoretical
consistency of the approach.

For the numerical calculation we employ the so-called Kahana-Ripka method
\cite{Kahana:1984be}, whose application to calculations of the nucleon EMT form
factors in the CQSM is briefly described in Ref.~\cite{accompanying-paper-I}.
The use of the proper-time regularization has the advantage
(over, e.g., the Pauli-Villars method \cite{Kubota:1999hx})
that it is possible to include explicitly chiral symmetry breaking effects due
to a finite current quark mass $m$ in the effective action (\ref{eff-theory}).

In Ref.~\cite{Goeke:2005fs} it was shown that it is possible to obtain
soliton solutions and compute nucleon masses for current quark masses
up to $m={\cal O}(700\,{\rm MeV})$ which corresponds to pion masses
up to $m_\pi={\cal O}(1.5\,{\rm GeV})$.
What provides a certain justification for the application of the model up such
large $m$ is the fact that the model {\sl formally} contains the
correct heavy quark limit result for the nucleon mass \cite{Goeke:2005fs}.
The proof that in the limit $m\to m_Q$, where $m_Q$ is the heavy quark mass,
the nucleon mass tends to $M_N\to N_cm_Q$ given in \cite{Goeke:2005fs} is
{\sl formal} because in this proof it was taken for granted that stable soliton
solutions do exist up to such large pion mass values.

Therefore, it was of importance in Ref.~\cite{Goeke:2005fs}  to demonstrate
numerically the existence of soliton solutions for large $m_\pi$
up to, at least, $m_\pi={\cal O}(1.5\,{\rm GeV})$.
These soliton solutions were found by using a standard iteration procedure
for the calculation of the self-consistent profile function $P_c(r)$
described in detail, e.g.\  in \cite{Christov:1995vm}.
Here, as a byproduct of our study of EMT form factors, we will be in the position
to provide an important and valuable cross check. Namely, do the pressures $p(r)$
computed with the respective large-$m_\pi$ soliton profiles really
satisfy the stability condition (\ref{Eq:stability})?
The answer is yes, see below Sec.~\ref{Sec-4d:stability}.

Before discussing the $m_\pi$-dependence of the densities
(\ref{Eq:density-energy},~\ref{Eq:density-pressure},~\ref{Eq:density-spin})
and the form factors
(\ref{Eq:M2-d1-model-comp},~\ref{Eq:d1-model-comp},~\ref{Eq:J-model-comp})
we have to establish which model parameters are allowed to vary and which
are kept fixed while $m_\pi$ is varied.
Here we shall use the choice of Ref.~\cite{Goeke:2005fs} to
keep $M=350\,{\rm MeV}$ and $f_\pi=93\,{\rm MeV}$ fixed. Then
the proper-time cutoff is adjusted such that for a given $m$ and $m_\pi$
one reproduces the physical value of $f_\pi$
($m$ and $m_\pi$ are related to each other in the effective theory
(\ref{eff-theory}) by a relation which for small $m$ corresponds
to the Gell-Mann---Oakes---Renner relation, see \cite{Goeke:2005fs}).

This way of parameter handling in the model was found to provide a good
description of lattice data on the variation of $M_N(m_\pi)$ with $m_\pi$,
once one takes into account the generic overestimate of the nucleon mass
in the soliton approach \cite{Pobylitsa:1992bk}.
However, the above way of parameter handling is just one possible choice and
other choices are possible as well.
An investigation whether other choices of parameter handling yield equally
satisfactory results will be presented elsewhere.

\newpage
\section{\boldmath The densities of the EMT}
\label{Sec-4:densities}

In this Section we shall focus our attention on the densities
(\ref{Eq:density-energy},~\ref{Eq:density-pressure},~\ref{Eq:density-spin})
which are interesting objects by themselves, before we discuss the form factors
(\ref{Eq:M2-d1-model-comp},~\ref{Eq:d1-model-comp},~\ref{Eq:J-model-comp})
in the next Section.
The study of the densities will enable us to address the question whether
the model provides a satisfactory description of the nucleon in (fictious)
worlds with pion masses up to $1.2\,{\rm GeV}$. As we shall see, the answer is yes.

\subsection{Energy density}
\label{Sec-4a:energy-density}

The energy density $\rho_E(r)$ is just $T^{00}(r)$ in the static EMT
(\ref{Def:static-EMT}), and is normalized as
$\int\di^3{\bf r}\,\rho_E(r,m_\pi)=M_N(m_\pi)$ for any $m_\pi$
where we explicitly indicate the pion mass dependence.
$M_N(m_\pi)$ as function of $m_\pi$ was studied in \cite{Goeke:2005fs}.

\begin{figure*}[b!]
\begin{tabular}{cc}
    \includegraphics[width=8.3cm]{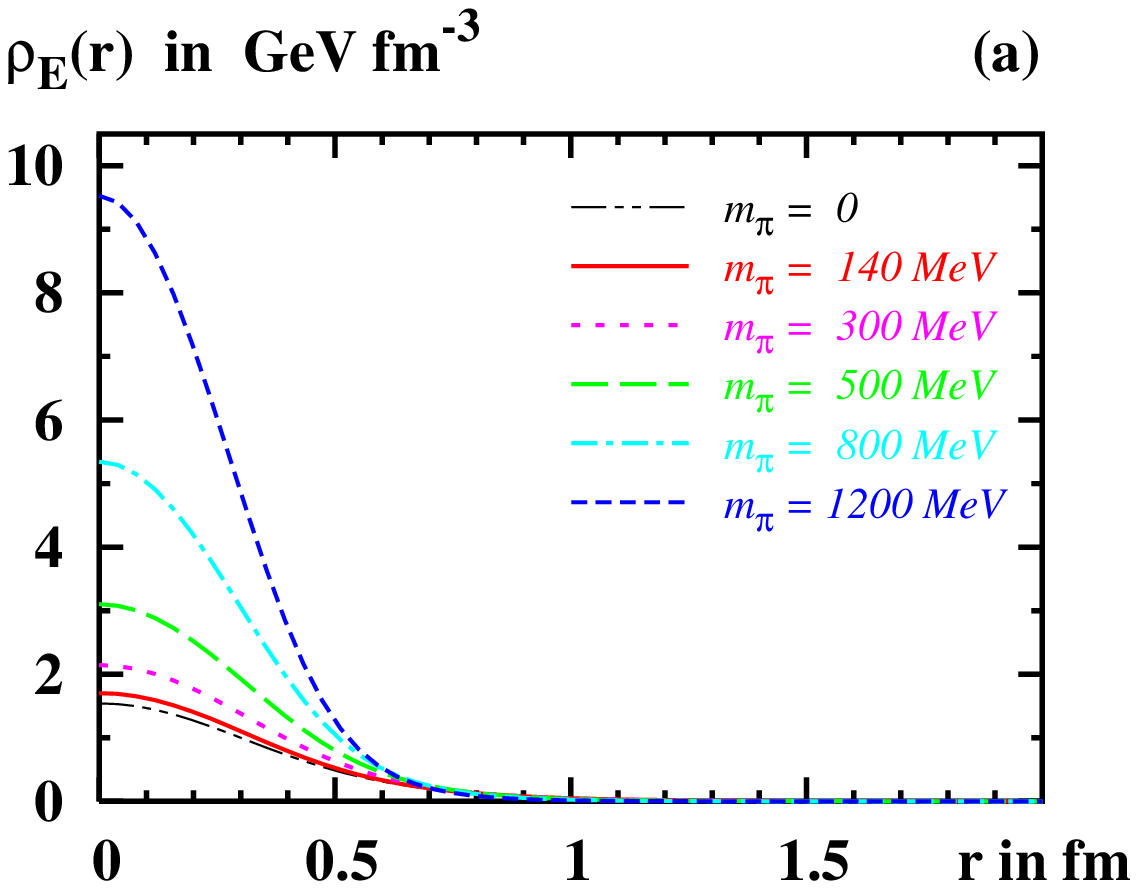}&
    \includegraphics[width=8.3cm]{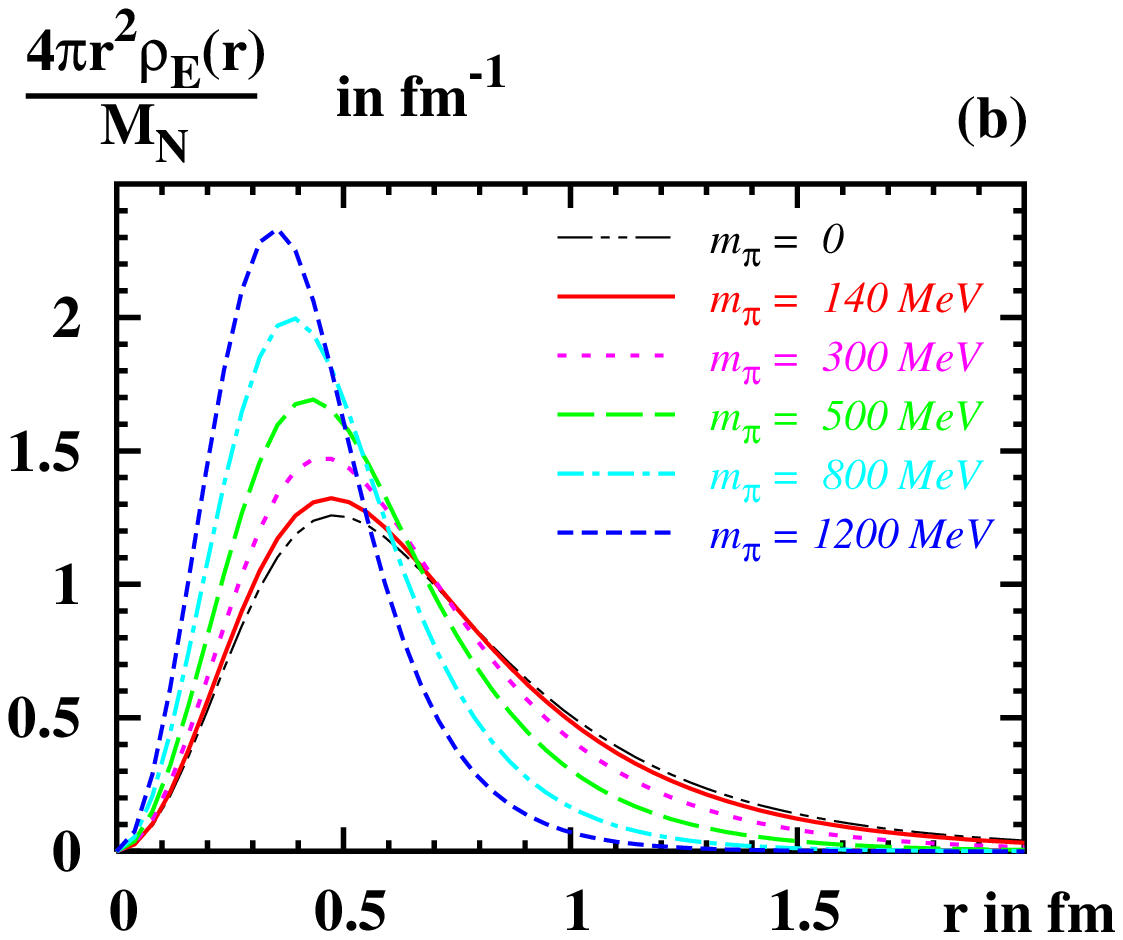}
\end{tabular}
    \caption{\label{Fig1-energy-density}
    \footnotesize\sl
    (a) The energy density $\rho_E(r)$ of the nucleon as function of $r$
    for different pion masses.
    (b) The normalized energy density $4\pi r^2\rho_E(r)/M_N$ as function
    for different pion masses. The curves are normalized such that one
    obtains unity upon integration over $r$.}
\begin{tabular}{cc}
    \includegraphics[width=8.3cm]{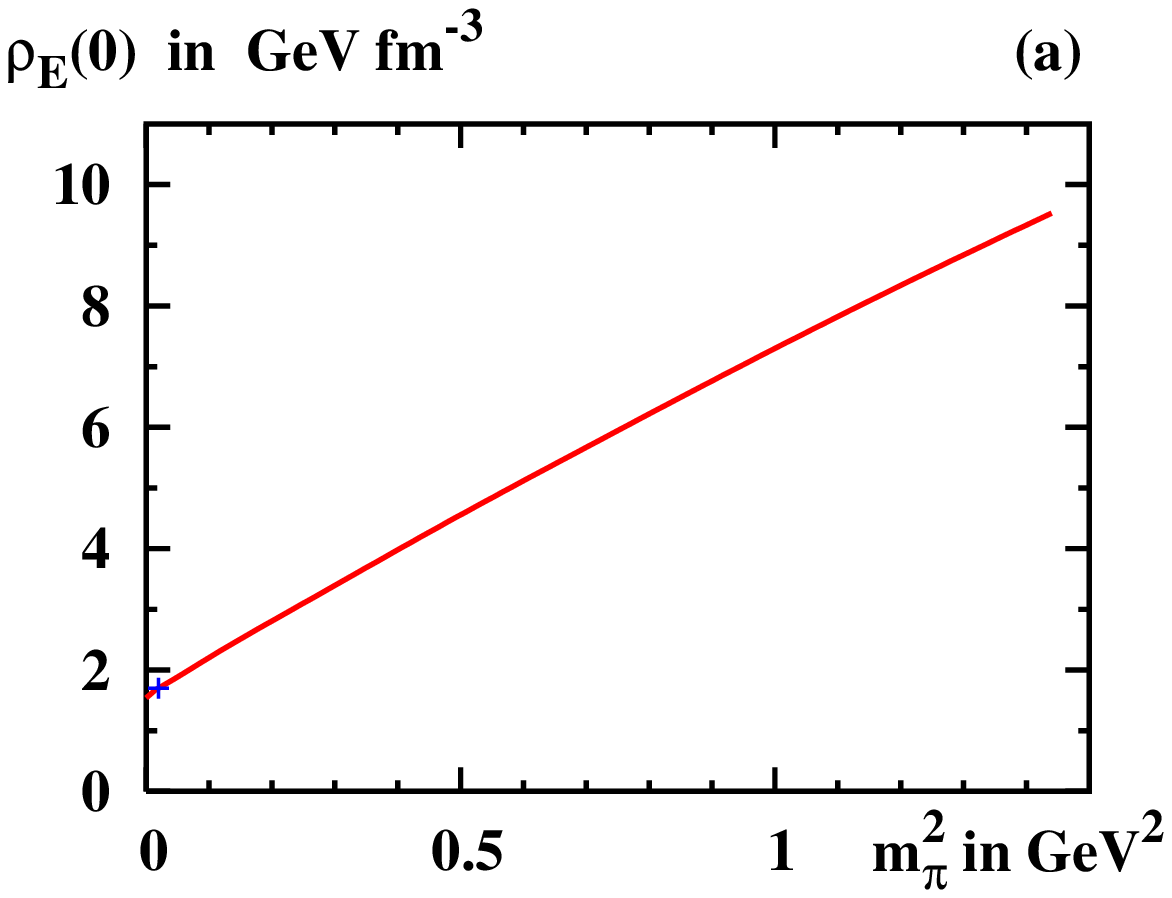}&
    \includegraphics[width=8.3cm]{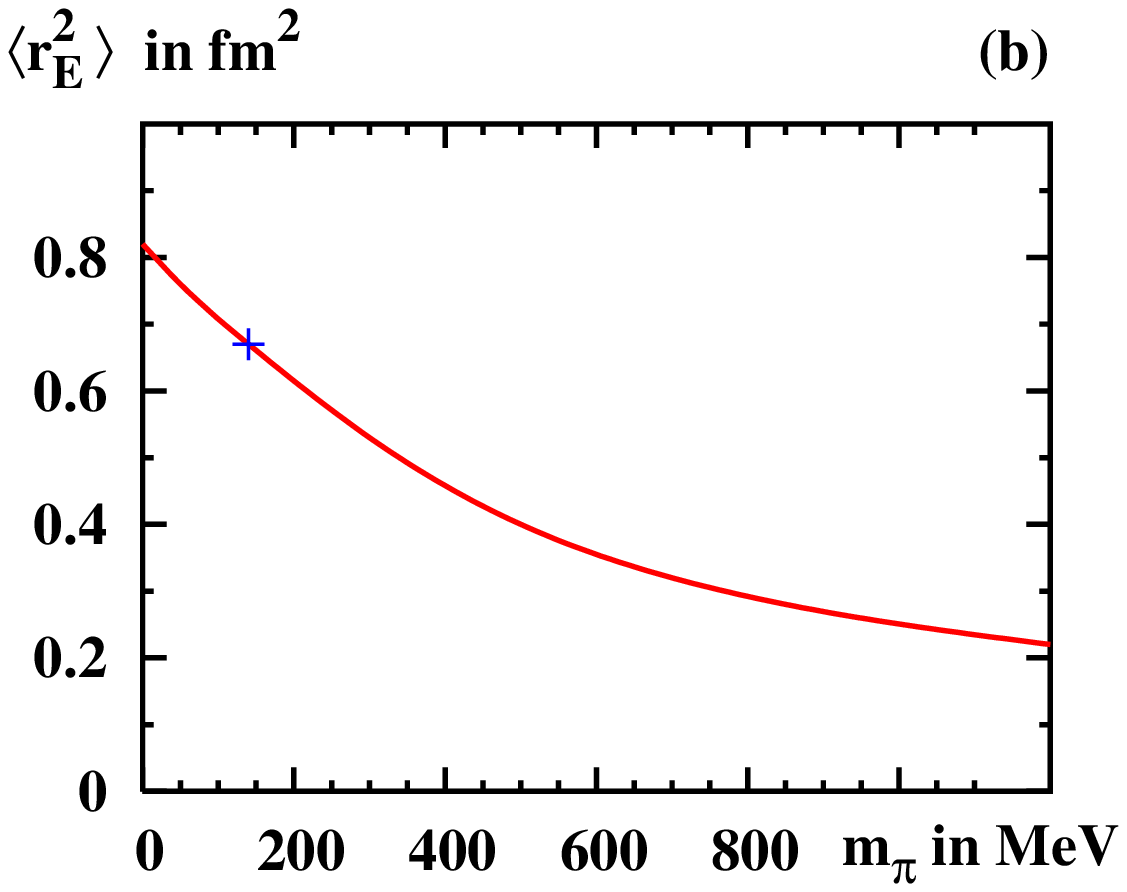}
\end{tabular}
    \caption{\label{Fig2-rho0-rE2}
    \footnotesize\sl
    The pion mass dependence of
    (a) the energy density $\rho_E(0)$ in the center of the nucleon, and
    (b) the mean square radius of the energy density $\la r_E^2\ra$
    defined in (\ref{Eq:def-energy-mean-square-radius}).
    The crosses indicate the physical point.}

 \end{figure*}

Fig.~\ref{Fig1-energy-density}a shows $\rho_E(r)$ as function
of $r$ for pion masses in the range $0 \le m_\pi \le 1.2\,{\rm GeV}$.
In the following we focus on the region $m_\pi \ge 140\,{\rm MeV}$, and
include only for completeness the results for $m_\pi \le 140\,{\rm MeV}$
discussed in detail in \cite{accompanying-paper-I}.

In the physical situation with $m_\pi=140\,{\rm MeV}$ the energy density in the center
of the nucleon is $1.7\,{\rm GeV\,fm}^{-3}$ or $3.0\times10^{15}\,{\rm g\,cm}^{-3}$.
This corresponds roughly to 13 times the equilibrium density of nuclear matter.
As $m_\pi$ increases $\rho_E(0)$ becomes larger and reaches
$\rho_E(0)=9.5\,{\rm GeV\,fm}^{-3}=17 \times 10^{15}\,{\rm g\,\,cm}^{-3}$
at $m_\pi =1.2\,{\rm GeV}$.
At the same time, with increasing $m_\pi$ the fall-off of $\rho_E(r)$ at large $r$
becomes stronger, as can be seen in Fig.~\ref{Fig1-energy-density}b.

These observations mean that with increasing $m_\pi$ the nucleon becomes ``smaller''.
To quantify this statement we consider the mean square radius of the energy density
defined as
\be\label{Eq:def-energy-mean-square-radius}
     \la r_E^2\ra =
    \frac{\int\di^3{\bf r}\;r^2\rho_E(r)}{\int\di^3{\bf r}\;\rho_E(r)}\;,\ee
which decreases with increasing $m_\pi$. The pion mass dependence of
$\rho_E(0)$ is shown in Fig.~\ref{Fig2-rho0-rE2}a, see also
Table~\ref{Table:square-radii-mpi} where many results are summarized.
We observe an approximately linear growth of $\rho_E(0)$ with $m_\pi^2$.
The pion mass dependence of $\la r_E^2\ra$ is shown in Fig.~\ref{Fig2-rho0-rE2}b,
see also Table~\ref{Table:square-radii-mpi}. Up to $m_\pi \lesssim 400\,{\rm MeV}$
we observe a roughly linear decrease of $\la r_E^2\ra$ with $m_\pi$ which proceeds
at a slower rate for  $m_\pi \gtrsim 400\,{\rm MeV}$.
(Throughout we choose a linear or quadratic in $m_\pi$ presentation of the
$m_\pi$-dependence of the quantities --- depending on which one is more convenient.)

The above observations can be intuitively understood. With increasing $m_\pi$
the range of the ``pion cloud'' decreases. This results in a less wide spread
nucleon. The above observations are also consistent with what one expects from
the heavy quark limit point of view. The heavier the constituents building up
a hadron, the smaller is the size of that hadron.
Thus, the model results for $\rho_E(r)$ are in agreement with what one expects
for increasing $m_\pi$.

We remark that, being a chiral model, the CQSM correctly describes the behaviour
of $\la r_E^2\ra $ in the chiral limit \cite{accompanying-paper-I}.

\subsection{\boldmath Angular momentum density}
\label{Sec-4b:spin-density}

\begin{figure*}[b!]
\begin{tabular}{cc}
    \includegraphics[width=8.3cm]{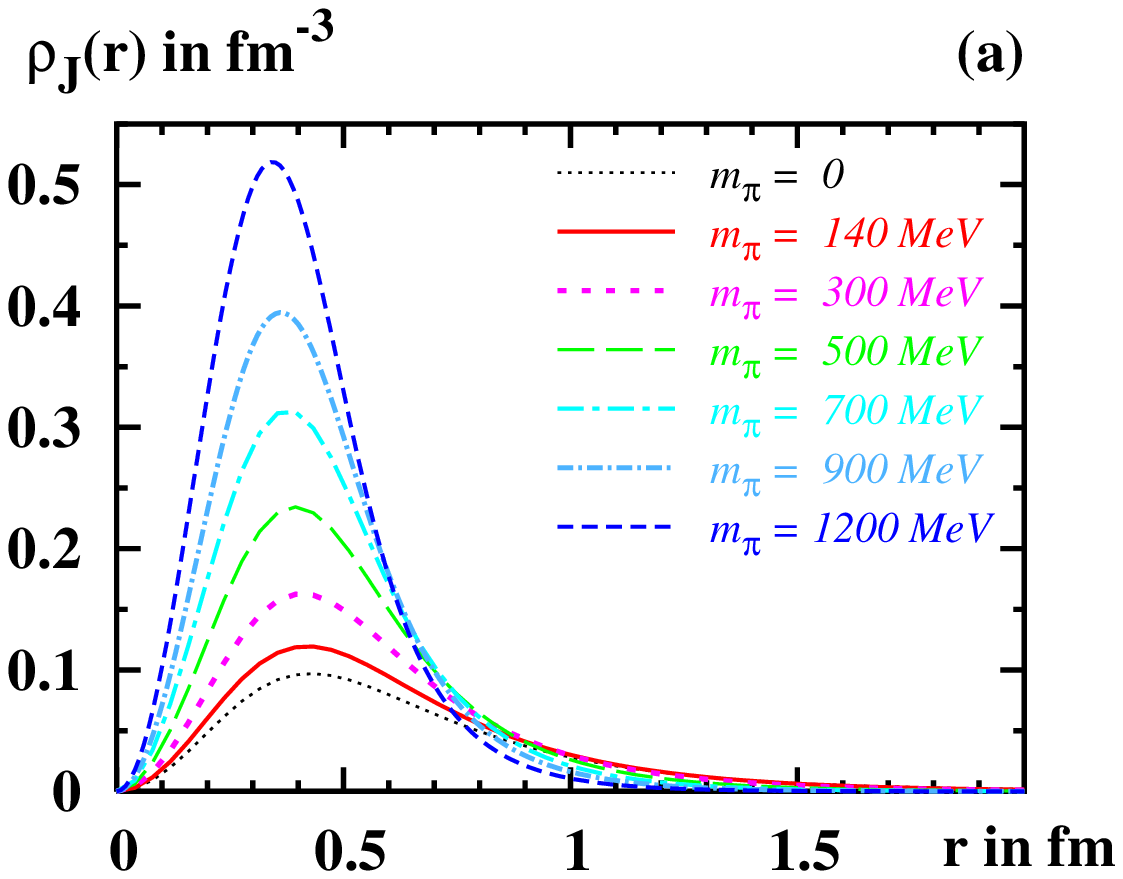}&
    \includegraphics[width=8.3cm]{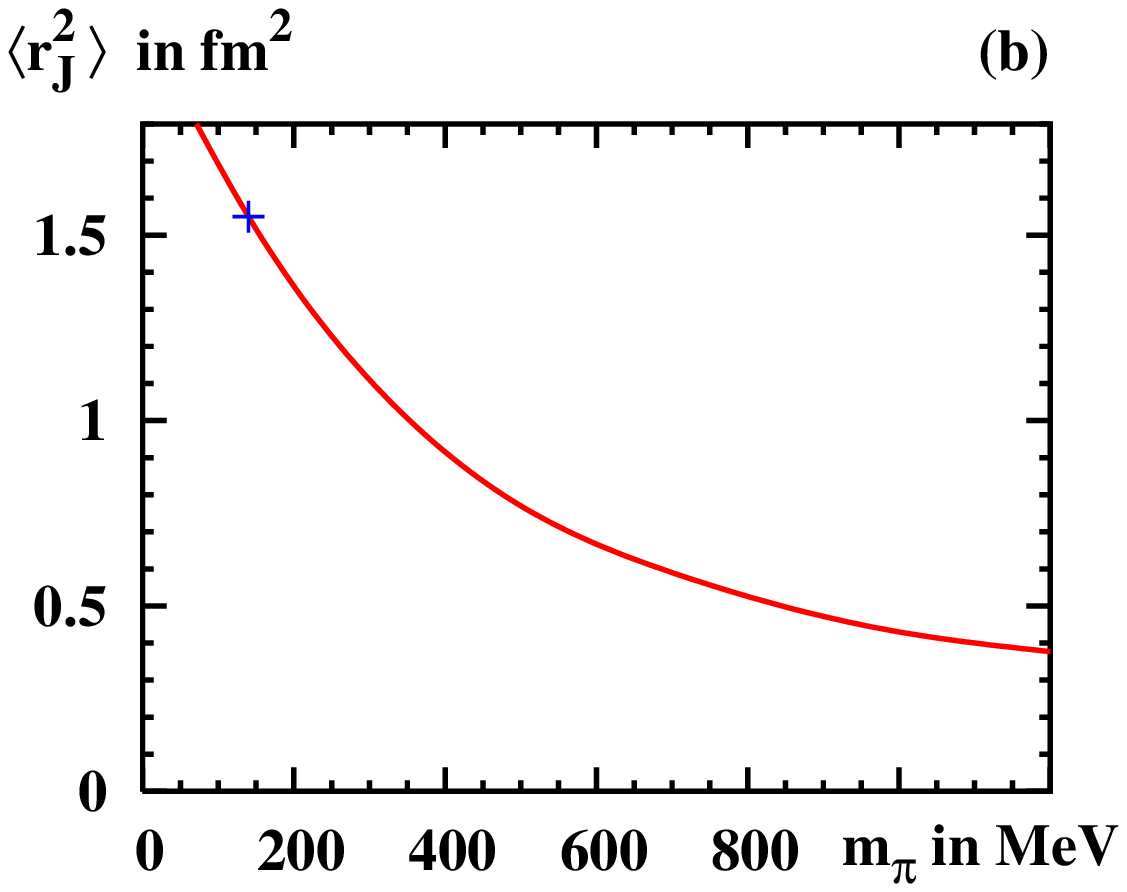}
\end{tabular}
    \caption{\label{Fig3-spin-density}
    \footnotesize\sl
    (a) The angular momentum density $\rho_J(r)$ as function of $r$ for
    different pion masses $m_\pi$.
    (b) The mean square radius $\la r_J^2\ra$ of the angular momentum density
    as function of $m_\pi$.}
\end{figure*}

The angular momentum density $\rho_J(r)$ is related to the $T^{0k}$ components
of the static EMT as $\rho_J(r) = \epsilon^{ijk}s_ix_jT_{0k}({\bf x})$.
Fig.~\ref{Fig3-spin-density}a shows $\rho_J(r)$ as function of $r$ for different
pion masses. For any $m_\pi$ we find that $\rho_J(r)\propto r^2$ at small $r$,
it reaches then a maximum around $r\simeq(0.3-0.4)\,{\rm fm}$ and goes slowly
to zero at large $r$.

As $m_\pi$ increases we observe that the density $\rho_J(r)$ becomes larger in
the small $r$ region at the price of decreasing in the region of larger $r$.
The increase in one and decrease in another region of $r$ (as $m_\pi$ is varied)
occurs in a precisely balanced way, because $\rho_J(r)$ ---
in contrast to the energy density --- is always normalized as
$\int\di^3{\bf r}\,\rho_J(r,m_\pi)=J_N=\frac12$, independently of $m_\pi$.
These observations can be understood within the picture of a rigidly rotating soliton
as follows. For large pion masses the ``matter'' inside the soliton is localized more
towards its center, as we have observed above, such that the inner region
of the soliton plays a more important role for its rotation.
As $m_\pi$ decreases, and hence the range of the pion cloud increases,
the energy density in the soliton becomes more strongly delocalized, and then the
``outer regions'' play a more and more important role for the rotation of the soliton.

These findings can be quantified by considering the mean square radius
$\la r_J^2\ra$ of the angular momentum density defined analogously to
(\ref{Eq:def-energy-mean-square-radius}).
Fig.~\ref{Fig3-spin-density}b shows $\la r_J^2\ra$ as function of $m_\pi$,
and we see that  $\la r_J^2\ra$ decreases with increasing $m_\pi$.
Notice that in the chiral limit $\rho_J(r) \propto 1/r^4$ at large $r$
such that  $\la r_J^2\ra$ diverges \cite{accompanying-paper-I}.
Our numerical results for $\la r_J^2\ra$ in Fig.~\ref{Fig3-spin-density}b
indicate this effect.

\newpage
\subsection{\boldmath Pressure and shear forces}
\label{Sec-4c:pressure-shear}

Next we turn to the discussion of the distributions of pressure and shear forces,
 $p(r)$ and $s(r)$, which are related to the $T^{ik}$ components of the static EMT.

\begin{figure*}[b!]
\vspace{-0.3cm}
\begin{tabular}{cc}
    \includegraphics[width=8.3cm]{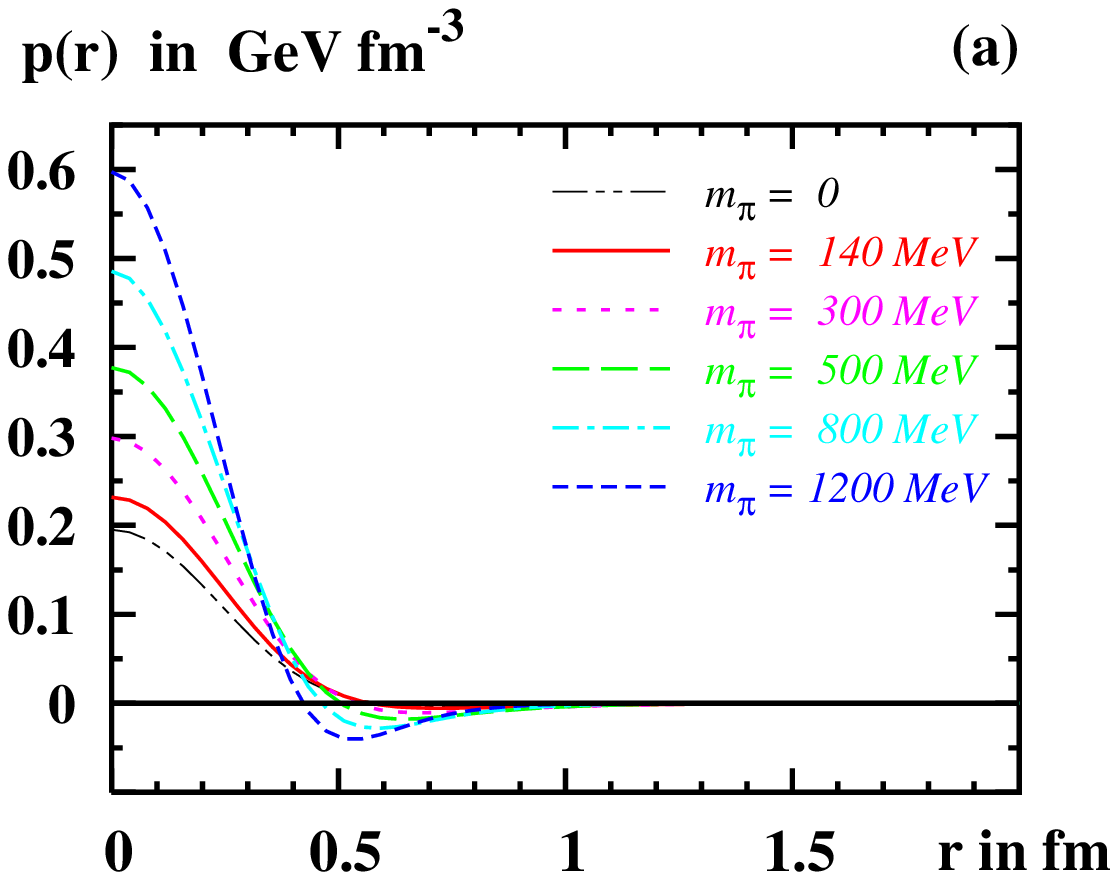}&
    \includegraphics[width=8.3cm]{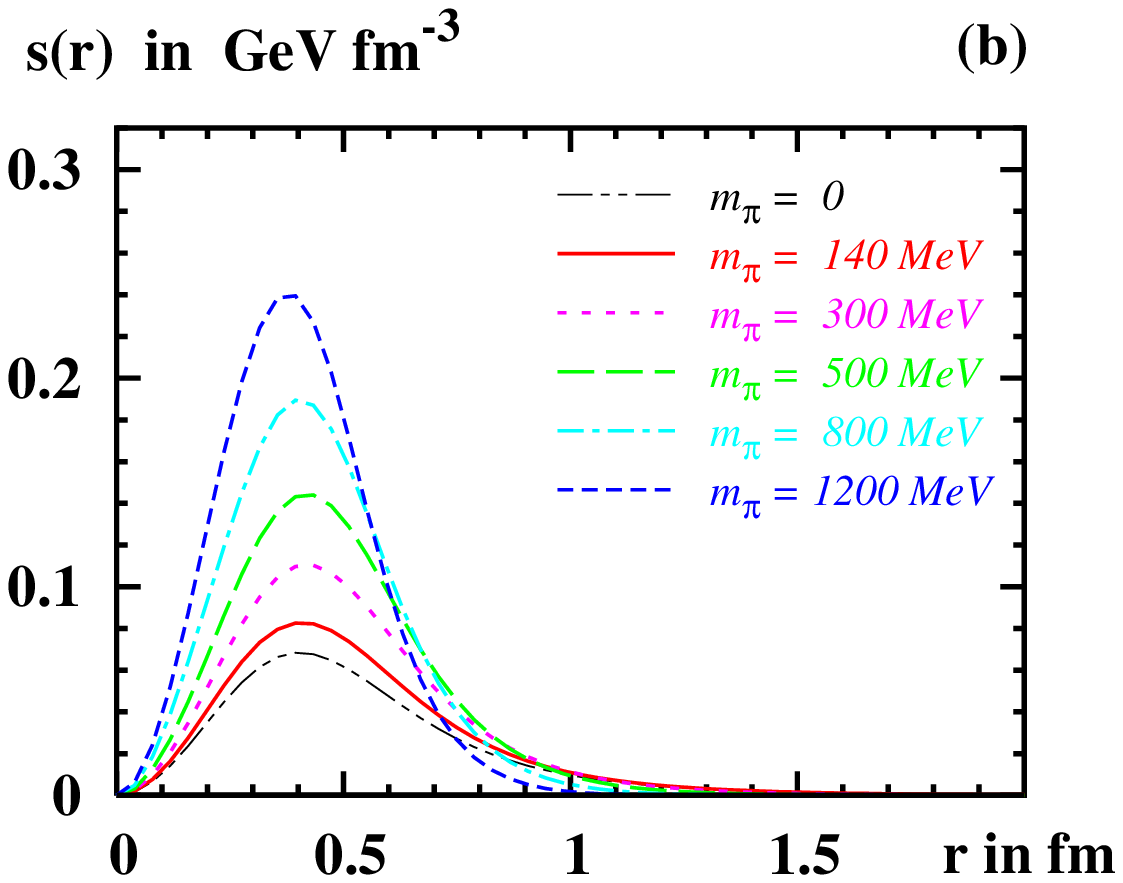}
\end{tabular}
    \caption{\label{Fig4-pressure+shear}
    \footnotesize\sl
    The distributions of (a) pressure $p(r)$ and (b) the shear forces $s(r)$
    as functions of $r$ for different pion masses.}
\begin{tabular}{cc}
    \includegraphics[width=8.3cm]{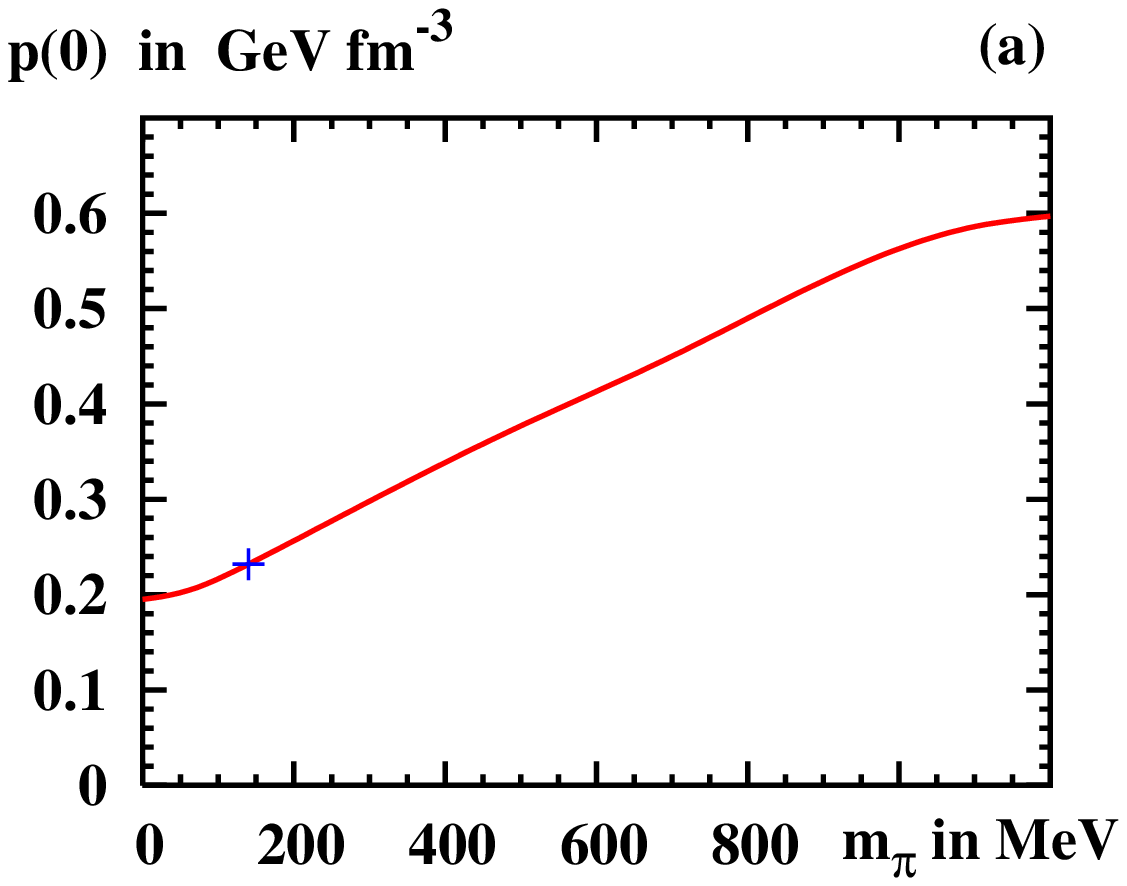}&
    \includegraphics[width=8.3cm]{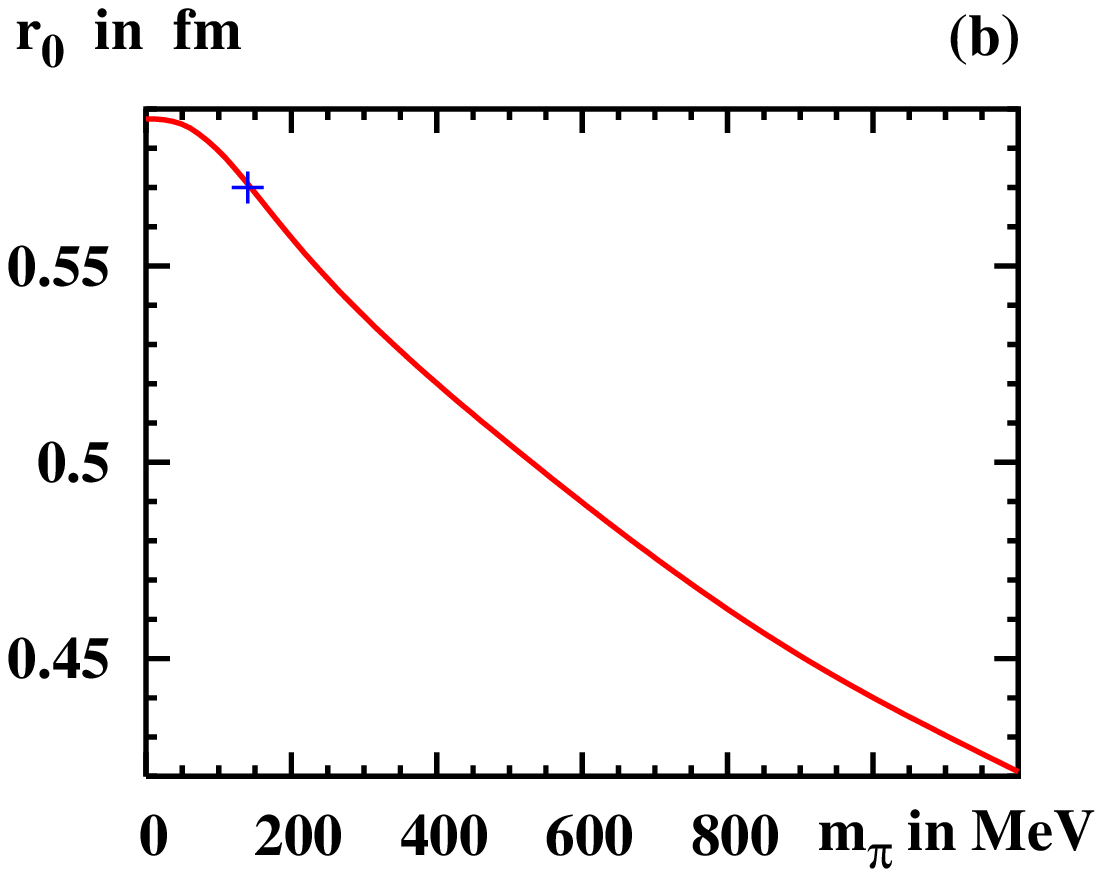}
\end{tabular}
    \caption{\label{Fig6-p0+r0}
    \footnotesize\sl
    The pion mass dependence of the pressure $p(0)$ in the center of the
    nucleon, and of the position $r_0$ at which $p(r)$ vanishes.}
\end{figure*}

Figs.~\ref{Fig4-pressure+shear}a and b show the distributions of pressure $p(r)$
and shear forces $s(r)$ as functions of $r$ for different $m_\pi$.
For all $m_\pi$ the distributions of pressure and shear forces exhibit the same
qualitative behaviour. The pressure takes at $r=0$ its global maximum,
decreases monotonically becoming zero at some point $r_0$ till reaching its global
minimum at some point $r_{p,\,\rm min}$, and decreases then monotonically tending
to zero but remaining always negative.
The distribution of shear forces is never negative. It starts at a zero value at $r=0$,
increases monotonically till reaching a global maximum at some point $r_{s,\rm max}$,
and decreases then monotonically tending to zero.

The positive sign of the pressure for $r<r_0$ corresponds to repulsion,
while the negative sign in the region $r > r_0$ means attraction.
This is intuitive because in the inner region we expect repulsion among
quarks due to the Pauli principle, while the attraction in the outer
region is an effect of the pion cloud which is responsible for binding
the quarks to form a nucleon \cite{accompanying-paper-I}.

With increasing $m_\pi$ the pressure $p(0)$ in the center of the nucleon increases.
At the same time also the absolute value of its (negative) minimum increases.
Also the maximum of $s(r)$ becomes larger with increasing $m_\pi$, while the
characteristic positions $r_0$, $r_{p,\,\rm min}$ and $r_{s,\,\rm max}$
move towards smaller $r$, see Fig.~\ref{Fig6-p0+r0} for $m_\pi$-dependence
of $p(0)$ and $r_0$.

These observations can be understood as follows.
The repulsive forces in the center of the nucleon increase as a response to the
higher density at larger $m_\pi$. At the same time the size of the nucleon decreases
requiring stronger binding forces --- and a ``movement'' of characteristic
length scales of $p(r)$ and $s(r)$ towards the center.
At any $m_\pi$ repulsive and attractive forces are precisely balanced due to
(\ref{Eq:stability}), see next Sec.~\ref{Sec-4d:stability}.

\subsection{Stability}
\label{Sec-4d:stability}

While the densities, $\rho_E(r)$ and $\rho_J(r)$, are normalized with
respect to $M_N$ and $J_N$, for the pressure $p(r)$ the corresponding analogon
is the stability criterion (\ref{Eq:stability}).
In Ref.~\cite{accompanying-paper-I} it was proven  analytically that
(\ref{Eq:stability}) is satisfied in the model --- provided one evaluates the
pressure with the self-consistent profile, i.e.\  with that profile which for a given
$m_\pi$ provides the true minimum of the soliton energy (\ref{Eq:soliton-energy}).

For given model parameters the soliton profiles are obtained by
means of an iteration procedure which is described in detail for
example in \cite{Christov:1995vm}. The profiles used here were
computed in Ref.~\cite{Goeke:2005fs} where a good convergence of
the iteration procedure was observed. However, what precisely
means that the convergence of the iteration was good? In other
words, how to test the quality of the numerical results? One
could, for example, slightly modify the obtained profiles and
check that they yield larger soliton masses than the respective
true self-consistent profile. But the probably most elegant method
is provided by the stability criterion (\ref{Eq:stability}). If,
and only if, we found the soliton profile which truely minimizes
the soliton energy (\ref{Eq:stability}), the pressure computed
with that profile will satisfy (\ref{Eq:stability}).

\begin{figure*}[t!]
\begin{tabular}{cc}
    \includegraphics[width=8.3cm]{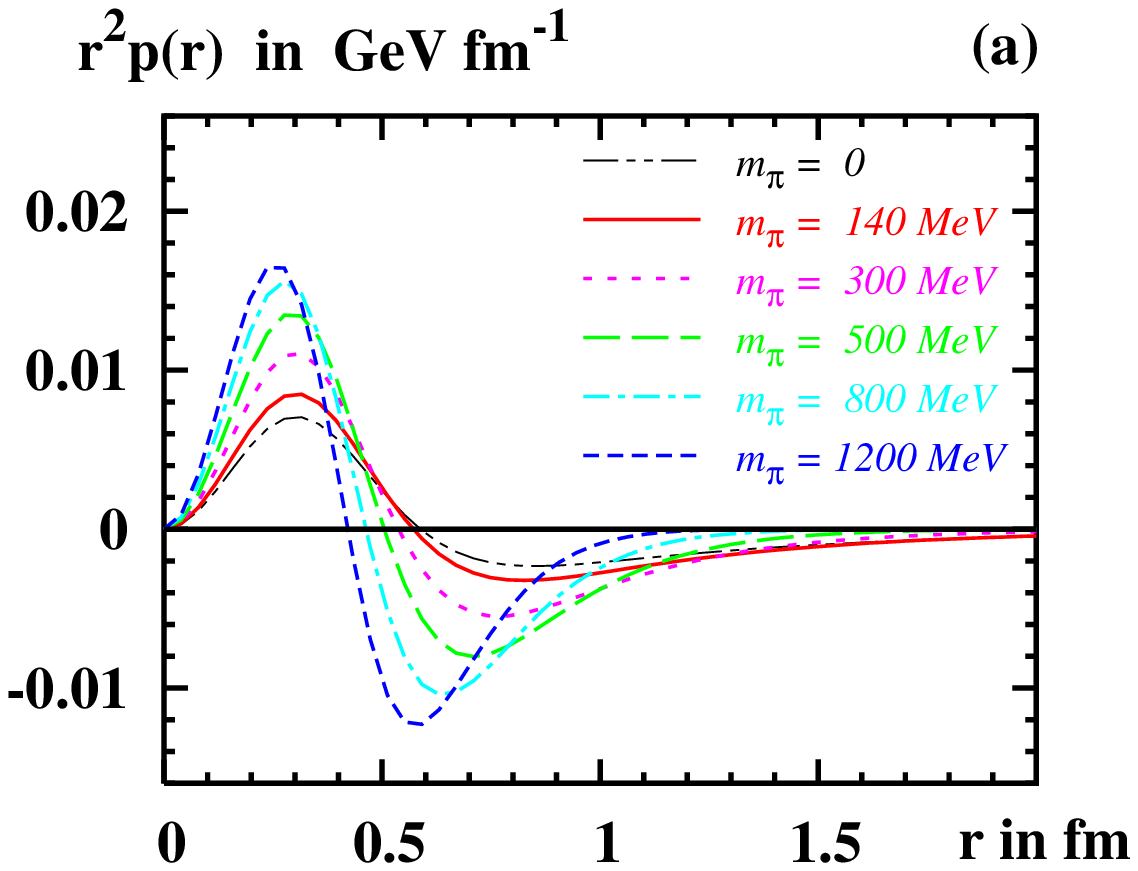}&
    \includegraphics[width=8.3cm]{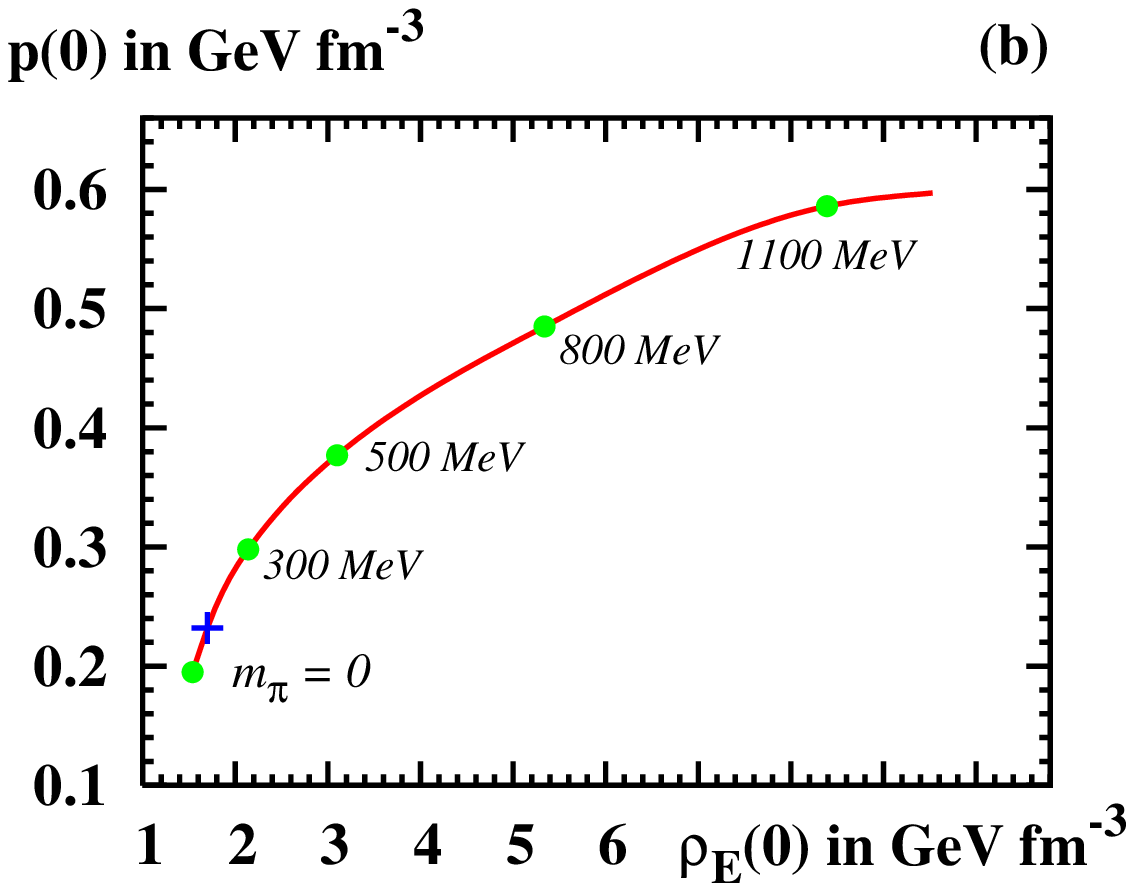}
\end{tabular}
    \caption{\label{Fig6-stability}
    \footnotesize\sl
    (a) $r^2p(r)$ as function of $r$ for different pion masses.
    The integrals over the regions where $r^2p(r)$ is respectively positive
    and negative cancel each other within numerical accuracy of better
    than $1\%$.
    This shows how the stability condition in (\ref{Eq:stability}) is
    satisfied, see text.
    (b) The pressure $p(0)$ as function of the energy density $\rho_E(0)$
    in the center of the nucleon. Some pion masses to which the respective
    values refer are indicated. The physical point is marked by the cross.}
\end{figure*}

One way to check to which numerical accuracy our results satisfy
(\ref{Eq:stability}) is as follows.
Let us consider $r^2p(r)$ as function of $r$, see Fig.~\ref{Fig6-stability}a,
and compute the integrals from $0$ to $r_0$ and from $r_0$ to $\infty$.\footnote{
    The numerical calculations are carried carried out in a finite spherical
    volume --- here of the size $D=12\,{\rm fm}$. For most quantities the
    densities decay fast enough at large $r$ such that it is sufficient
    to integrate up to $r=D$. This is what we did in Eq.~(\ref{Eq:pressure-02}).
    However, for certain quantities given by integrals over the densities
    weighted by a higher power of $r$ the integrands may happen not to be
    negligibly small at large $r$ in particular for small
    $m_\pi\lesssim 140\,{\rm MeV}$.
    Then it is necessary to explore the analytically known large-$r$
    asymptotics of the densities, see \cite{accompanying-paper-I},
    and to include the contribution
    of the regions $r>D$ not covered in the numerical calculation.
    Examples of such quantities are $d_1$ or $\la r_J^2\ra$. The latter
    is divergent in the chiral limit, see Sec.~\ref{Sec-4b:spin-density}.}
We obtain
\be\label{Eq:pressure-02}
    \int\limits_0^{r_0} \di r\;r^2p(r) = \cases{
     2.614\,{\rm MeV} \cr
     3.737\,{\rm MeV} \cr
     3.856\,{\rm MeV}, }\;\;\;
    \int\limits_{r_0}^\infty\di r\;r^2p(r) = \cases{
    -2.630\,{\rm MeV} \cr
    -3.748\,{\rm MeV} \cr
    -3.861\,{\rm MeV}, }
    \;\;\;
    \mbox{i.e.}
    \;\;\;
    \frac{|\int_0^\infty\di r\;r^2p(r)|}{\int_0^\infty\di r\;r^2|p(r)|}
     = \cases{
    0.31\, \% & for $m_\pi=140\,{\rm MeV}$ \cr
    0.15\, \% & for $m_\pi=500\,{\rm MeV}$ \cr
    0.07\, \% & for $m_\pi=1200\,{\rm MeV}$,}
\ee
and see that the stability criterion is satisfied to within a satisfactory numerical
accuracy. 

Finally, we may test another sort of stability. We may ask the question
how do pressure and energy density in the center of nucleon depend on
each other for varying $m_\pi$. In fact, with $\rho_E(r,m_\pi)$
and $p(r,m_\pi)$ at hand, we may eliminate $m_\pi$ at $r=0$ and express
$p(0)$ as function of $\rho_E(0)$. This is shown in Fig.~\ref{Fig6-stability}b
which demonstrates how the center of the nucleon responds to changes of $m_\pi$.
Understanding the center of the nucleon for a moment as a ``medium which is subject
to variations of the external parameter $m_\pi$'' we observe that for any $m_\pi$
we have $\frac{\partial p(\rho_E)}{\partial\rho_E}>0$. This is a criterion for
stability of a system which must respond with an increase of pressure if the
density is increased.

\newpage
\section{\boldmath Results for the form factors}
\label{Sec-5:form-factors}

\begin{figure*}[b!]
\begin{tabular}{ccc}
  \hspace{-0.5cm}
  \includegraphics[width=6cm]{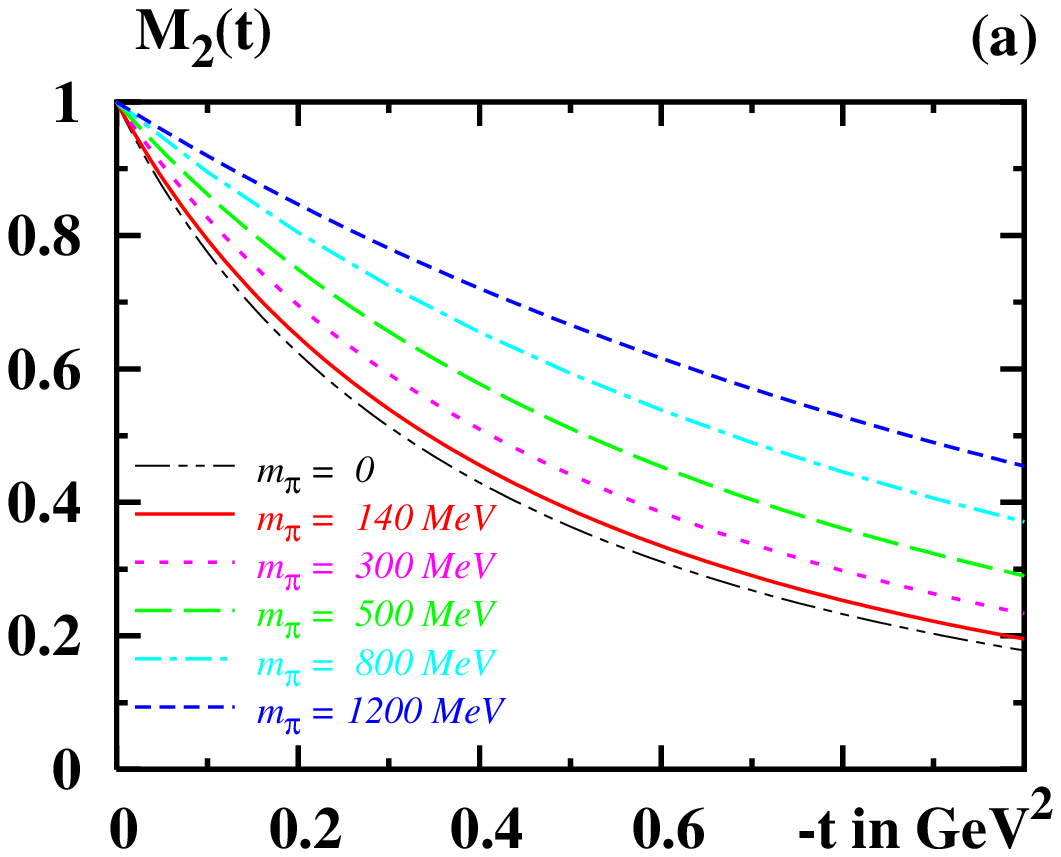}&
  \includegraphics[width=6cm]{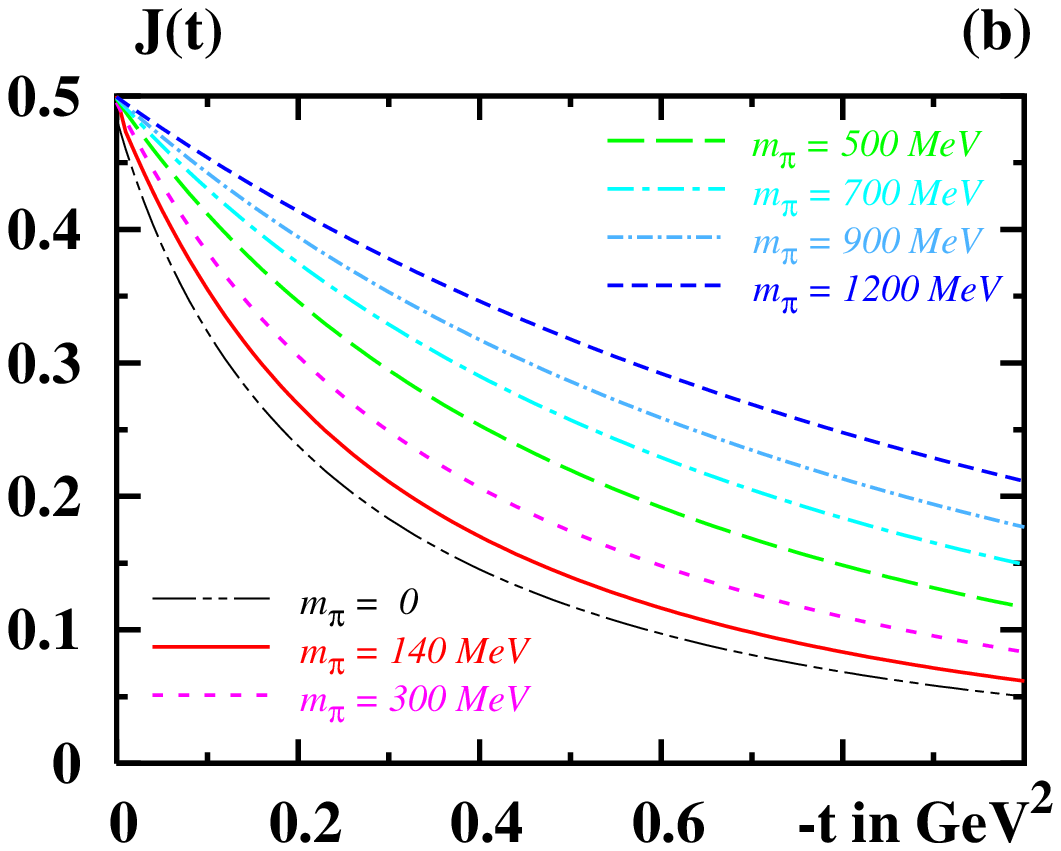}&
  \includegraphics[width=6cm]{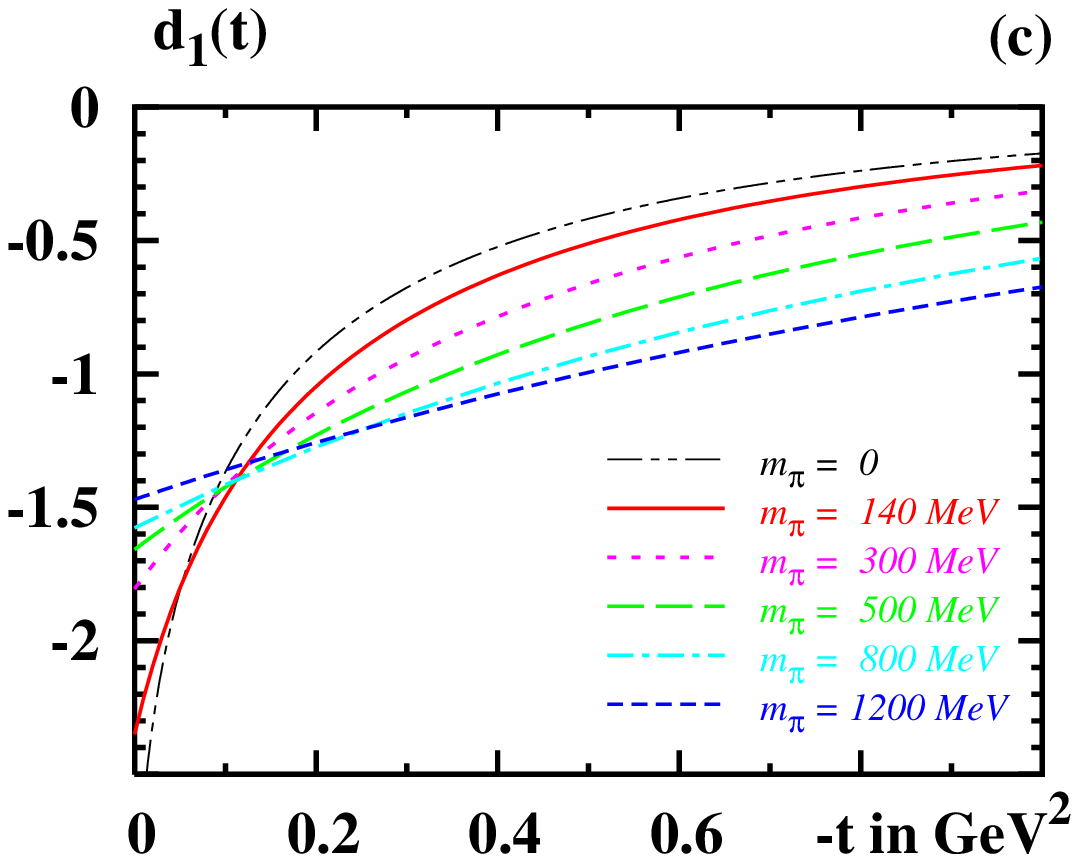}
\end{tabular}
    \caption{\label{Fig7-ffs}
    \footnotesize\sl
    The form factors of the energy momentum tensor
    $M_2(t)$, $J(t)$ and $d_1(t)$ as functions of $t$ for
    different pion masses.
    In the chiral limit $J(t)$ and $d_1(t)$ exhibit infinitely steep
    slopes at $t=0$, and $d_1(0)$ takes the value $-3.46$ which does not
    fit on the scale in Fig.~\ref{Fig7-ffs}c.}
\end{figure*}

Fig.~\ref{Fig7-ffs} shows the form factors of the EMT as functions of $t$ for
$|t|\le 1\,{\rm GeV}^2$ for different $m_\pi$. All EMT form factors
(with the exception of $J(t)$ and $d_1(t)$ in the chiral limit, see below)
can be well approximated by dipole fits of the kind
\be\label{Eq:dipol}
    F(t) = \frac{F(0)}{(1-t/M_{\rm dip}^2)^2}\;.
\ee
It is instructive to compare within the model the EMT form factors to the
electromagnetic form factors --- for example to the electric form factor of
the proton $G_E(t)$ \cite{Christov:1995hr}. Interestingly, $J(t)$ and $G_E(t)$
show a similar $t$-dependence. But $M_2(t)$ falls off with increasing $|t|$
slower than $G_E(t)$, while $d_1(t)$ exhibits a faster fall off,
see \cite{accompanying-paper-I} for more details.

\begin{figure*}[t!]
\begin{tabular}{cc}
  \hspace{-0.5cm}
  \includegraphics[width=8cm]{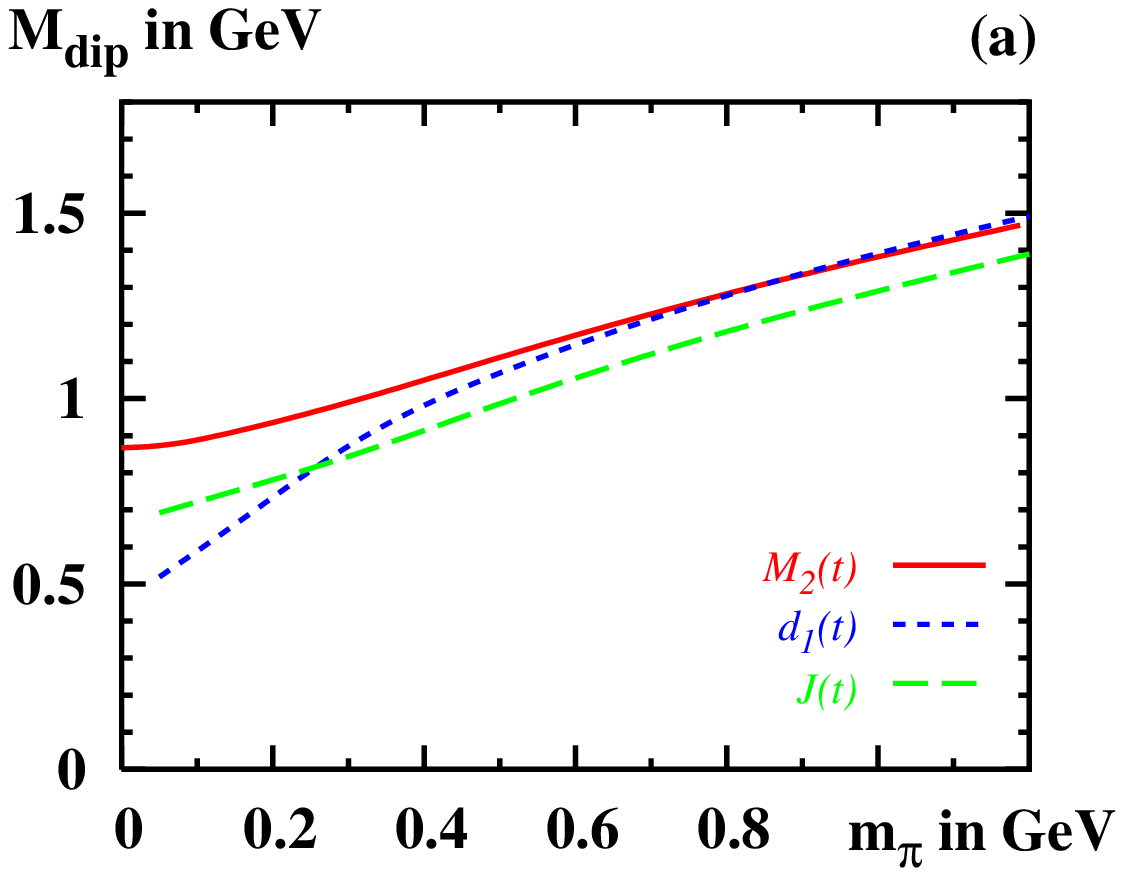} &
  \includegraphics[width=8cm]{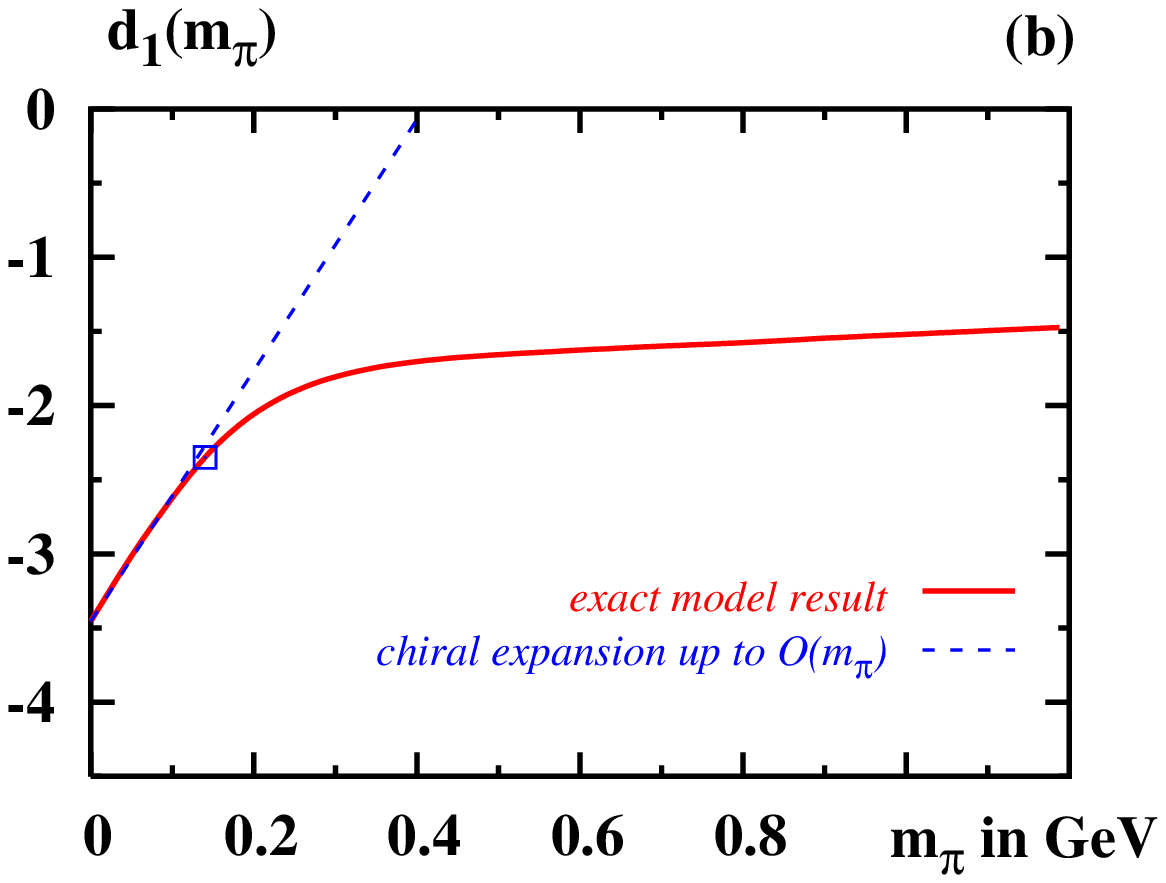}
\end{tabular}
    \caption{\label{Fig8-d1-dipol-mpi}
        \footnotesize\sl
    (a) The dipole masses as defined in (\ref{Eq:dipol})
    for the form factors $M_2(t)$, $d_1(t)$ and $J(t)$
    from the CQSM vs.\ $m_\pi$. Notice that
    $d_1(t)$ and $J(t)$ cannot be approximated by dipoles
    for $m_\pi=0$.
    (b) The constant $d_1$ as function of $m_\pi$ at low scale.
    Solid line: The exact result from the CQSM.
    Dotted line: The chiral expansion of $d_1(m_\pi)$ to
    linear order in $m_\pi$ according to (\ref{Eq:d1-vs-mpi}).
    The square marks the physical point.}
\end{figure*}

The dipole masses of the different form factors exhibit different $m_\pi$-dependences,
see Fig.~\ref{Fig8-d1-dipol-mpi}a and Table~\ref{Table:square-radii-mpi}.
For all form factors the dipole masses increase with increasing $m_\pi$.
It is an interesting observation that the dipole masses of $M_2(t)$ and $J(t)$
exhibit for $m_\pi\gtrsim 140\,{\rm MeV}$ to a good approximation a linear
dependence on $m_\pi$. But the $m_\pi$-dependence of the dipole mass of $d_1(t)$
follows a different pattern.
We shall comment more on that in Sec.~\ref{Sec-7:lattice-lessons}

That the dipole approximation for $J(t)$ and $d_1(t)$ fails in the chiral limit,
is due to fact that the slopes of  $J(t)$ and $d_1(t)$ at $t=0$ diverge in this
limit \cite{accompanying-paper-I}.
For $J(t)$ this is clear because its derivative at $t=0$
is related to the mean square radius of the angular momentum density
as $J^\prime(0) = \frac16 \la r_J^2\ra$, and $\la r_J^2\ra$ diverges
for $m_\pi\to0$, see Sect.~\ref{Sec-4b:spin-density}.

The slope of $d_1(t)$ at $t=0$ becomes infinitely steep in the chiral limit
because it is related as
$d_1^\prime(0) = \frac{M_N}{16}\,\int\di^3{\bf r}\;r^4p(r)$ to the pressure
which behaves as $p(r)\propto\frac{1}{r^6}$ at large $r$ in the chiral limit.
More precisely, $d_1^\prime(t)\propto 1/\sqrt{-t}$ at small $t$ in the chiral limit.
For small but non-zero $m_\pi$ the derivative $d_1^\prime(0)$ exists and is
proportional to $1/m_\pi$. These results hold both in the CQSM
\cite{accompanying-paper-I} and in $\chi$PT \cite{Belitsky:2002jp}.
The derivative of the form factor $M_2(t)$ at $t=0$ is finite for any $m_\pi$.

$M_2(t)$ and $J(t)$ are normalized at $t=0$ for any $m_\pi$ as
$M_2(0)=2J(0)=1$ \cite{accompanying-paper-I}.
In the CQSM these constraints are consistent for they mean that
entire momentum and spin of the nucleon are carried by quark degrees of freedom.
The numerical results satisfy these constraints within a numerical
accuracy of better than $1\,\%$, see Figs.~\ref{Fig7-ffs}a and \ref{Fig7-ffs}b.

In contrast, no principle fixes the normalization of the form factor
$d_1(t)$ at $t=0$ neither in the model nor in QCD.
For all $m_\pi$ we find $d_1=d_1(0)<0$. This condition has been
conjectured to be dictated by stability requirements \cite{accompanying-paper-I}.
Fig.~\ref{Fig8-d1-dipol-mpi}b shows the $m_\pi$-dependence of $d_1$
which is rather strong. This is due to the fact that $d_1$ receives a
large leading non-analytic contribution proportional to $m_\pi$.
(The ``non-analyticity'' refers to the current quark mass $m\propto m_\pi^2$.)
The chiral expansion of $d_1$ reads
\be\label{Eq:d1-vs-mpi}
    d_1(m_\pi)=\;\kringel{d_1}\;+\;k\;\frac{5\,g_A^2M_N}{64\,\pi\,f_\pi^2}\;m_\pi
    + \dots
\ee
where $\kringel{d_1}$ denotes the chiral limit value of $d_1$,
and the dots indicate subleading terms in the chiral limit.
Since the limits $N_c\to\infty$ and $m_\pi\to 0$ do not commute
\cite{Dashen:1993jt,Cohen:1992uy} one has in Eq.~(\ref{Eq:d1-vs-mpi})
$k=1$ for finite $N_c$ \cite{Belitsky:2002jp} and
$k=3$ in the large-$N_c$ limit \cite{accompanying-paper-I}.
The latter corresponds to the situation in the CQSM.

It is interesting to observe that the leading non-analytic term in the chiral
expansion of $d_1$ in Eq.~(\ref{Eq:d1-vs-mpi}) dominates the chiral behaviour
of $d_1$ up to the physical point, see Fig.~\ref{Fig8-d1-dipol-mpi}b.
But for larger $m_\pi$ higher orders in the chiral expansion become important,
and change the qualitative $m_\pi$-behaviour of $d_1(m_\pi)$.
We shall come back to this point in Sec.~\ref{Sec-7:lattice-lessons}.

\begin{wrapfigure}[14]{R}{5.8cm}
\hspace{-0.7cm} \includegraphics[width=6.2cm]{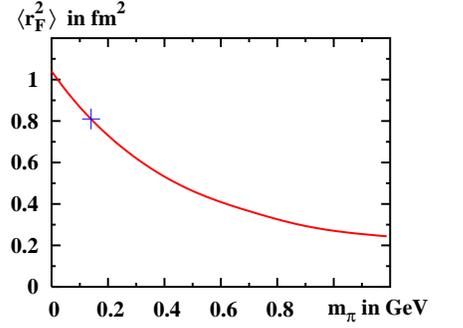}
    \caption{\label{Fig9-rF2}
    \footnotesize\sl
    The mean square radius $\la r_F^2\ra$ of the trace
    of the total EMT operator vs.\  $m_\pi$.}
\end{wrapfigure}

Finally, we discuss $m_\pi$-dependence of the mean square radius $\la r_F^2\ra$ of
the trace of the total EMT operator given due to the trace anomaly \cite{Adler:1976zt}
by
\be\label{Eq:ff-of-EMT-trace-1}
    \hat{T}_{\!\mu}^{\,\mu} \equiv \frac{\beta}{2g}\;F^{\mu\nu}F_{\mu\nu}
    +(1+\gamma_m)\sum_am_a\bar\psi_a\psi_a \;.\ee
Let $F(t)$ denote the form factor of the operator (\ref{Eq:ff-of-EMT-trace-1}) which
is normalized as $F(0)=1$. Its slope at $t=0$ defines $\la r_F^2\ra = 6F^\prime(0)$
which can be related as
\be\label{Eq:ff-of-EMT-trace-2}
\la r_F^2\ra = \la r_E^2\ra -\frac{12\,d_1}{5M_N^2}
\ee
to $\la r_E^2\ra$ and $d_1$, see \cite{accompanying-paper-I}.
Fig.~\ref{Fig9-rF2} shows how  $\la r_F^2\ra$ depends on the pion mass.
In the chiral limit $\la r_F^2\ra$ is the mean square radius of the operator
$F^{\mu\nu}F_{\mu\nu}$ and its large value there is in contrast to what is known
about the mean square radii of other gluonic operators \cite{Braun:1992jp}.

%
\begingroup
\squeezetable
\begin{table}[b!]
\vspace{-1cm}
    \caption{\footnotesize\sl
    \label{Table:square-radii-mpi}
  The pion mass dependence of different quantities computed in the CQSM
  using proper time regularization:
  the energy density in the center of the nucleon $\rho_E(0)$,
  the mean square radii $\la r_E^2\ra$ and $\la r_J^2\ra$ ,
  the pressure $p(0)$ in the center of the nucleon,
  the position $r_0$ of the zero of the pressure defined as $p(r_0)=0$,
  the constant $d_1$,
  the dipole masses of the form factors $M_2(t)$, $J(t)$ and $d_1(t)$,
  and the mean square radius $\la r_F^2\ra$ of the trace of the EMT operator.
  In the chiral limit $J(t)$ and $d_1(t)$ have infinitely steep slopes at $t=0$,
  see text. In these cases dipole fits do not provide useful approximations and
  are undefined (labelled by ``undef.'' in the Table).
  The results for $m_\pi\le 140\,{\rm MeV}$ from Ref.~\cite{accompanying-paper-I}
  are included for completeness.}
\vspace{0.2cm}
    \begin{ruledtabular}
    \begin{tabular}{ccccccccccccc}
    \\
&
$m_\pi$     &
$\rho_E(0)$ &
$\la r_E^2\ra$  &
$\la r_J^2\ra$  &
$p(0)$      &
$r_0$       &
$d_1$       &
\multicolumn{3}{l}{dipole masses $M_{\rm dip}$ in GeV for} &
$\la r_F^2\ra$  &
$\;\;\;\;\;$ \cr
&
${\rm MeV}$ &
${\rm GeV}/{\rm fm}^3$&
${\rm fm}^2$&
${\rm fm}^2$&
${\rm GeV}/{\rm fm}^3$&
${\rm fm}$ &
&
$M_2(t)$ &
$J(t)$   &
$d_1(t)$ &
${\rm fm}^2$ &
\cr
    \\
    \hline
\cr
&0   & 1.54& 0.82& $\infty$& 0.195& 0.59& -3.46& 0.867& undef.& undef.& 1.04&\\
&50  & 1.57& 0.76& 1.88    & 0.202& 0.59& -3.01& 0.873& 0.692 & 0.519 & 0.95&\\
&140 & 1.70& 0.67& 1.55    & 0.232& 0.57& -2.35& 0.906& 0.745 & 0.646 & 0.81&\\
&300 & 2.14& 0.53& 1.11    & 0.298& 0.54& -1.81& 0.990& 0.844 & 0.872 & 0.62&\\
&500 & 3.10& 0.40& 0.77    & 0.377& 0.51& -1.66& 1.111& 0.986 & 1.069 & 0.46&\\
&700 & 4.50& 0.32& 0.59    & 0.450& 0.49& -1.60& 1.228& 1.120 & 1.214 & 0.37&\\
&900 & 6.86& 0.26& 0.48    & 0.553& 0.46& -1.55& 1.334& 1.237 & 1.337 & 0.29&\\
&1200& 9.53& 0.22& 0.38    & 0.597& 0.42& -1.47& 1.473& 1.390 & 1.492 & 0.24&\\
\cr
\end{tabular}
\end{ruledtabular}
\end{table}
\endgroup
%

\newpage

\section{\boldmath Comparison to lattice results}
\label{Sec-6:lattice}

It is instructive to compare the results for the form factors of the EMT to lattice
QCD data
\cite{Mathur:1999uf,Gadiyak:2001fe,Hagler:2003jd,Gockeler:2003jf,Negele:2004iu}.
Presently, this offers actually the only available test for the model results.
For the comparison it is necessary to evolve the model results from a low initial
scale $\mu_0\sim 0.6\,{\rm GeV}$ to typically $\mu\sim 2\,{\rm GeV}$ in the lattice
calculations which we shall do to leading logarithmic accuracy.
Under evolution the quark (flavour-singlet) form factors mix with the corresponding
gluon form factors. We set the latter to zero at the initial scale.

In this context it is worthwhile recalling that in early parameterizations
of unpolarized parton distributions $f_1^a(x)$, the gluon (and sea quark)
distribution(s) and thus $M_2^G$ were assumed to be zero at a low initial scale
\cite{Gluck:1977ah}.
One success of these approaches was that they were able to explain the 
observation $M_2^Q\approx 0.5$ at $\mu^2 \sim 5\,{\rm GeV}^2$.
I.e.\  starting with $M_2^Q = 1$ and $M_2^G=0$ at a low scale, which is
the situation in the CQSM, it is possible to reproduce the observation
$M_2^Q\approx 0.5$ at several ${\rm GeV}^2$.
However, with the advent of more data (especially at low $x$) it became clear
\cite{Gluck:1994uf} that non-zero gluon (and see quark) distributions are required
already at low initial scales. Modern parameterizations performed at low scales
require a sizeable gluon distribution and $M_2^G\approx 0.3$ \cite{Gluck:1998xa}.
This is not in disagreement with the instanton picture where
twist-2 gluon operators are suppressed with respect to quark operators
by the instanton packing fraction which is numerically of order $30\,\%$
\cite{Diakonov:1995qy}.
Thus in some sense the phenomenologically required ``portion of gluons'' is
within the accuracy of the model \cite{Diakonov:1998ze}.
With these remarks in mind we conclude that the CQSM result $M_2^Q=1$ at the low
scale of the model is in agreement with phenomenology --- within the accuracy
of this model. 

Next, before we compare the results from the model to those from lattice QCD,
let us confront the lattice results
\cite{Mathur:1999uf,Gadiyak:2001fe,Hagler:2003jd,Gockeler:2003jf,Negele:2004iu}
with predictions from the large-$N_c$ limit. In this limit one has
for $|t|\ll M_N^2$ independently of the scale \cite{Goeke:2001tz}
\ba\label{Eq:large-Nc-ABCq}
    \underbrace{ (A^u+A^d)(t) }_{{\cal O}(N_c^0)}    \gg
    \underbrace{|(A^u-A^d)(t)|}_{{\cal O}(N_c^{-1})} \; , \;\;\; 
    \underbrace{|(B^u-B^d)(t)|}_{{\cal O}(N_c)}      \gg
    \underbrace{|(B^u+B^d)(t)|}_{{\cal O}(N_c^0)}    \; , \;\;\; 
    \underbrace{|(C^u+C^d)(t)|}_{{\cal O}(N_c^2)}    \gg
    \underbrace{|(C^u-C^d)(t)|}_{{\cal O}(N_c)} \;,  \ea
cf.\  App.~\ref{App:Alternative-definition} for the explanation of the
notation. For completeness let us quote that the gluon form factors satisfy
\be\label{Eq:large-Nc-ABCg}
    A^G(t)={\cal O}(N_c^0)\, , \;\;\;
    B^G(t)={\cal O}(N_c^0)\, , \;\;\;
    C^G(t)={\cal O}(N_c^2)\, , \ee
which is the same large-$N_c$ behaviour as the corresponding
quark-flavour-singlet form factors.

Remarkably, although in the real world $N_c=3$ does not seem to be large, nevertheless
lattice data
\cite{Mathur:1999uf,Gadiyak:2001fe,Hagler:2003jd,Gockeler:2003jf,Negele:2004iu}
reflect
the large-$N_c$ flavour dependence of the quark form factors (\ref{Eq:large-Nc-ABCq}).
In fact, large-$N_c$ relations of the type
(\ref{Eq:large-Nc-ABCq},~\ref{Eq:large-Nc-ABCg}) are observed to
be satisfied in phenomenology \cite{Efremov:2000ar} and serve within
their range of applicability as useful guidelines \cite{Pobylitsa:2003ty}.
The soliton approach is justified in the large-$N_c$ limit
\cite{Witten:1979kh} and satisfies general large-$N_c$ relations
of the type (\ref{Eq:large-Nc-ABCq}). The observation that the lattice results
\cite{Mathur:1999uf,Gadiyak:2001fe,Hagler:2003jd,Gockeler:2003jf,Negele:2004iu}
are compatible with (\ref{Eq:large-Nc-ABCq}) is therefore an encouraging
prerequisite for our study.

\begin{figure*}[t!]
\begin{tabular}{cc}
  \hspace{-0.5cm}
  \includegraphics[width=6cm]{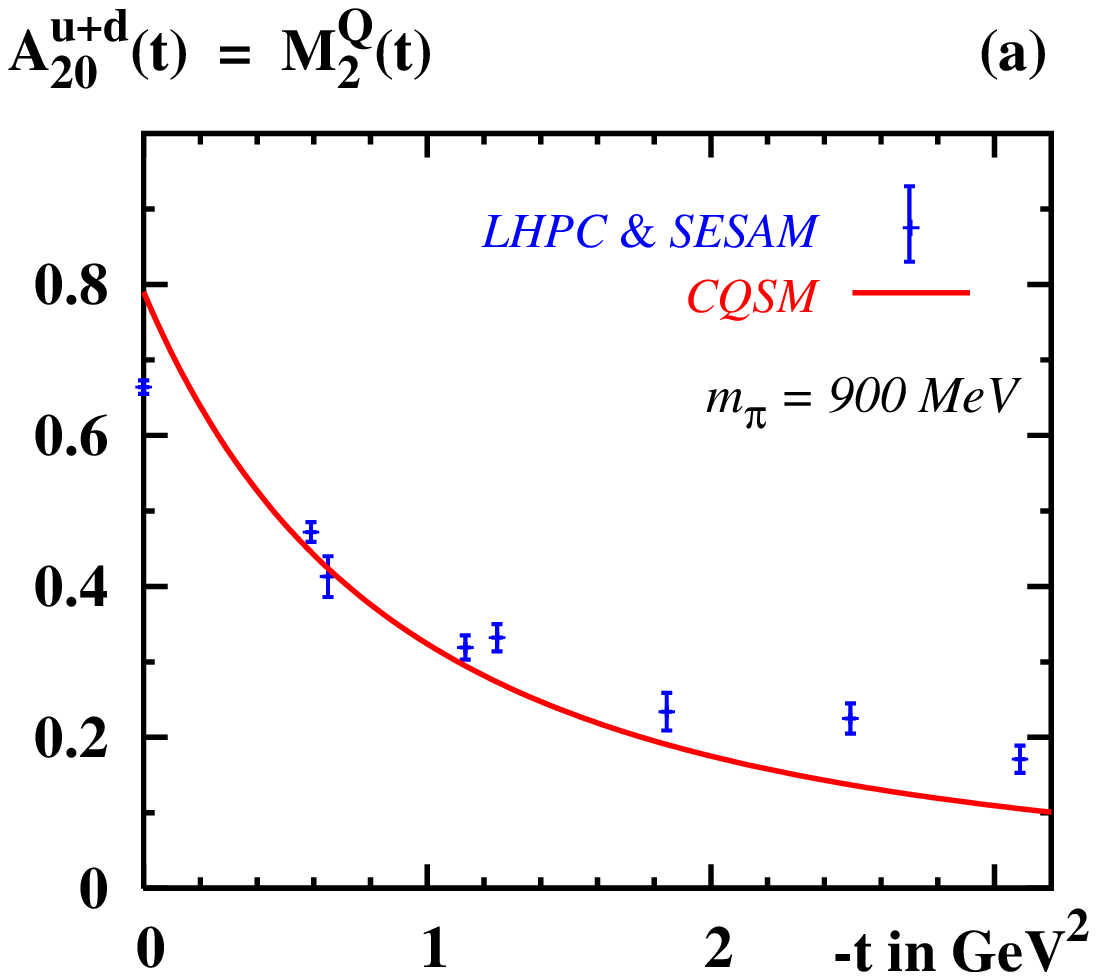}&
  \includegraphics[width=6cm]{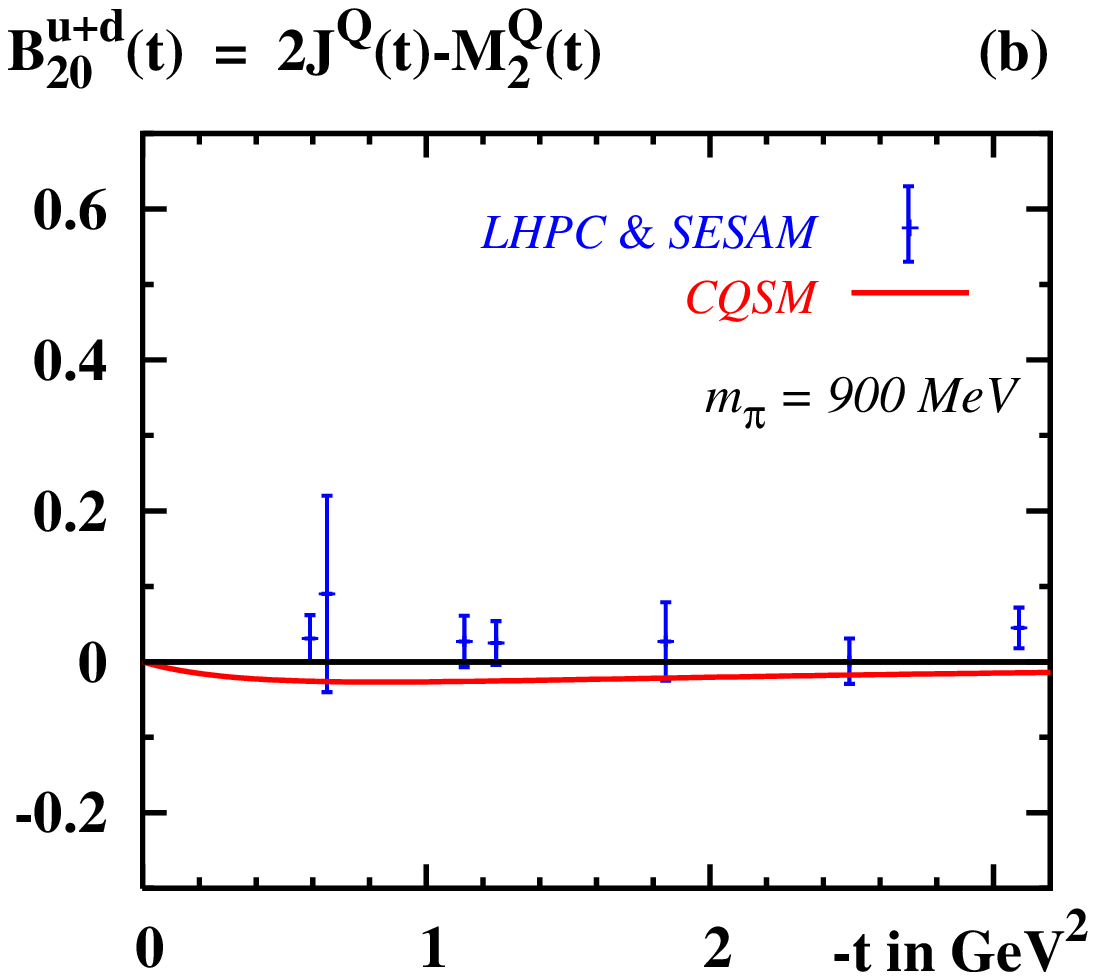}
\end{tabular}
        \caption{\label{Fig10-lattice-mpi-900}
    \footnotesize\sl
    The form factors of the energy momentum tensor
    $A_{20}^{u+d}(t)\equiv A^Q(t) = M_2^Q(t)$ and
    $B_{20}^{u+d}(t)\equiv B^Q(t) = 2J^Q(t)-M_2^Q(t)$ as functions of $t$
    at the scale $\mu=2\,{\rm GeV}$ in a world with pion mass
    $m_\pi=900\,{\rm GeV}$. (See App.~\ref{App:Alternative-definition}
    for the discussion of different notations.)
    The lattice data points are by the LHPC and SESAM Collaborations
    \cite{Hagler:2003jd}. The solid curves are the results from the
    CQSM obtained here.}
\begin{tabular}{ccc}
  \hspace{-0.5cm}
  \includegraphics[width=6cm]{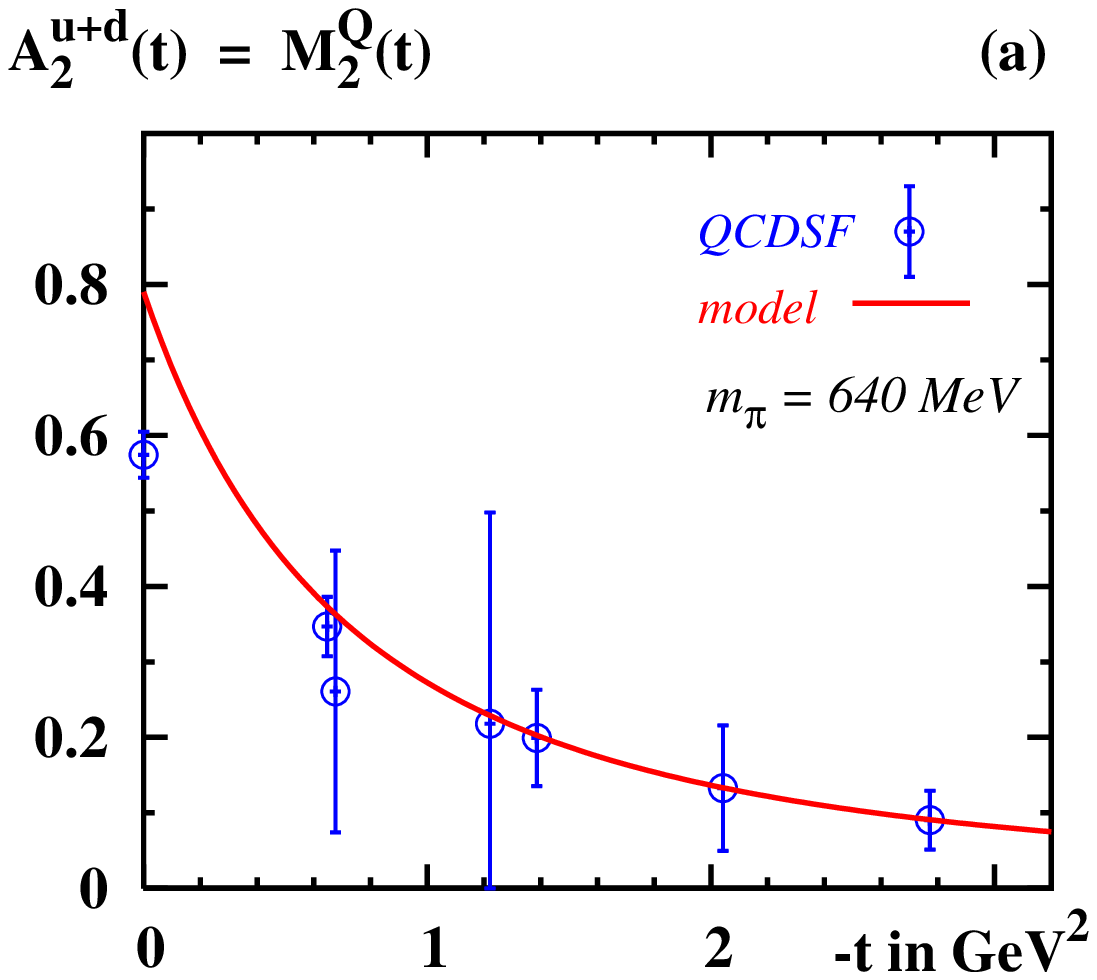}&
  \includegraphics[width=6cm]{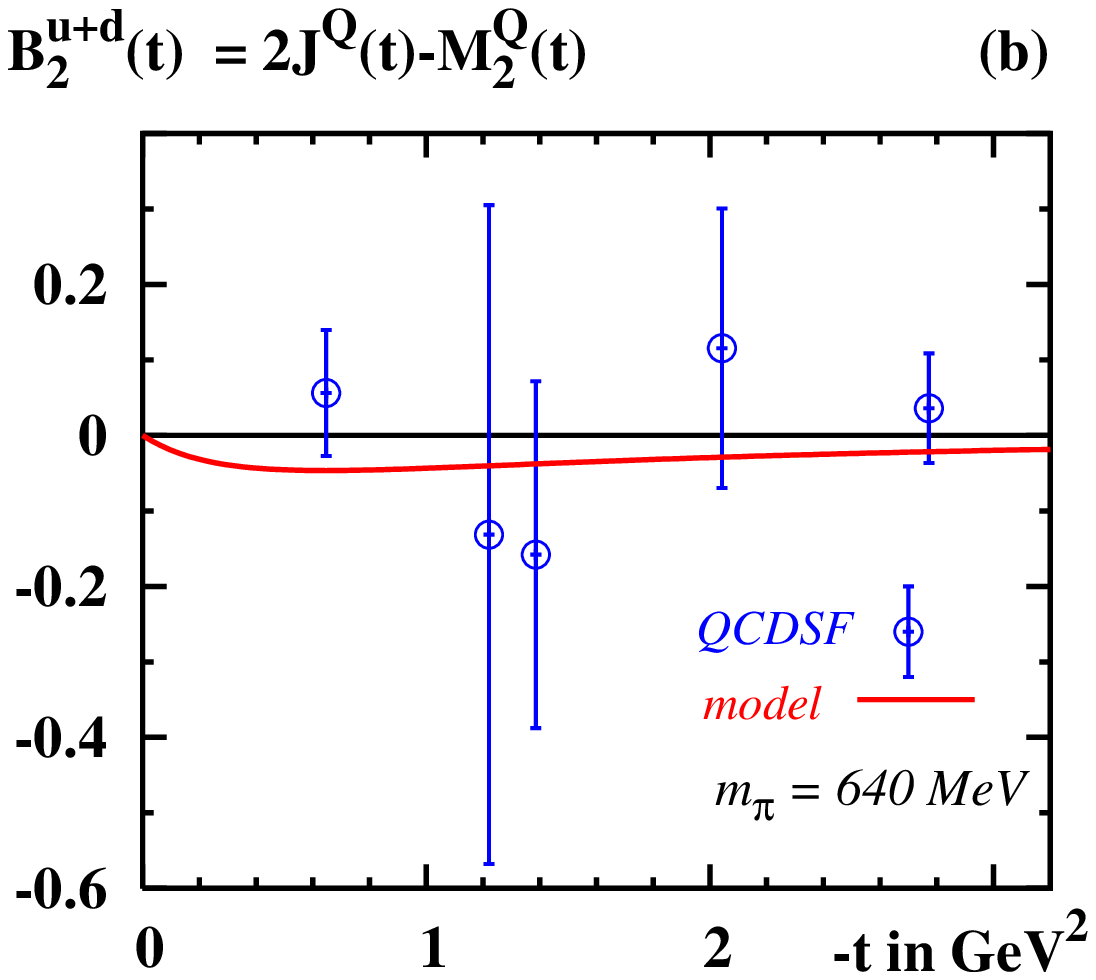}&
  \includegraphics[width=6cm]{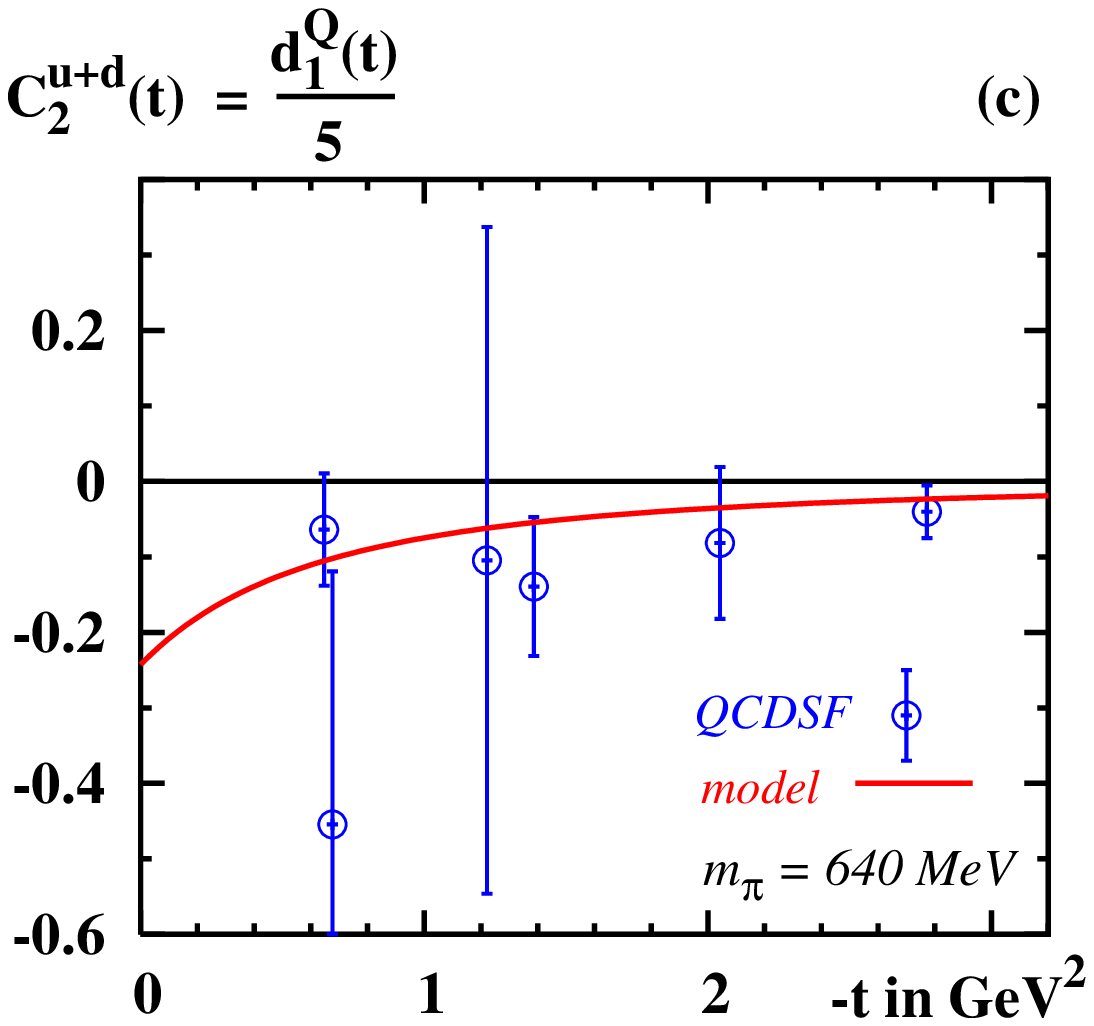}
\end{tabular}
        \caption{\label{Fig11-lattice-mpi-640}
    \footnotesize\sl
    The form factors of the energy momentum tensor
    $A_{2}^{u+d}(t)\equiv A^Q(t) = M_2^Q(t)$ and
    $B_{2}^{u+d}(t)\equiv B^Q(t) = 2J^Q(t)-M_2^Q(t)$ and
    $C_{2}^{u+d}(t)\equiv C^Q(t) = \frac15 d_1^Q(t)$ as functions of $t$
    at the scale $\mu=2\,{\rm GeV}$ in a world with pion mass
    $m_\pi=640\,{\rm GeV}$. (For notations see App.~\ref{App:Alternative-definition}.)
    The lattice data points are from the QCDSF Collaboration
    \cite{Gockeler:2003jf}. The solid curves are the results from the
    CQSM obtained here.}
\end{figure*}

Let us first compare the model results to the lattice data computed by the LHPC and
SESAM Collaborations \cite{Hagler:2003jd}. There unquenched SESAM 
Wilson configurations on a $16^3\times 32$ lattice at $\beta=5.6$ with
$\kappa=0.1560$ were used.
This corresponds to $m_\pi=(896\pm 6)\,{\rm MeV}$ and a lattice spacing of
$a=0.098\,{\rm fm}$ with physical units fixed by extrapolating the nucleon mass.
The form factor $A_{20}^{u+d}(t)=A^Q(t)\equiv M_2^Q(t)$ was computed omitting
disconnected diagrams at a scale $\mu=2\,{\rm GeV}$ for $0\le|t|\le 3.1\,{\rm GeV}^2$.
(The different notations are discussed in App.~\ref{App:Alternative-definition}.)
The lattice data for $A^Q(t)$, which can be fit to the dipole form,
are shown in Fig.~\ref{Fig10-lattice-mpi-900}a.
In Fig.~\ref{Fig10-lattice-mpi-900}a we also show the CQSM results evolved to
the same scale for $|t|\lesssim 1\,{\rm GeV}^2$.
We observe that the model results agree with the lattice data \cite{Hagler:2003jd}
to within $15\,\%$ which is within the accuracy to which the CQSM results typically
agree with phenomenology.

Fig.~\ref{Fig10-lattice-mpi-900}b shows the form factor
$B_{20}^{u+d}(t)=B^Q(t)\equiv 2J^Q(t)-M_2^Q(t)$
from Ref.~\cite{Hagler:2003jd} which was computed in the range
$-t\in [0.59,3.1]\,{\rm GeV}$ and was found consistent with zero within the
statistical accuracy of the simulation. Also in the CQSM we find $B^Q(t)$
close to zero --- in reasonable agreement with the lattice data, see
Fig.~\ref{Fig10-lattice-mpi-900}b. Notice that in the model, at the low
scale, we have $B^Q(0)=0$ since $M_2^Q(0)=2J^Q(0)$ (and equal unity), see
Secs~\ref{Sec-4a:energy-density} and \ref{Sec-4b:spin-density}. This implies
a vanishing quark contribution to the "gravitomagnetic moment" of the nucleon
\cite{Ossmann:2004bp} conjectured in \cite{Teryaev:1998iw}.
The smallness of $B^Q(t)$ implies that to a good approximation
$2J^Q(t)\approx M_2^Q(t)$. We shall come back to this point below.

Next, let us compare to the quenched lattice results by the QCDSF Collaboration
\cite{Gockeler:2003jf} which were obtained using non-perturbatively
${\cal O}(a)$ improved Wilson fermions on a $16^3\times 32$ lattice at $\beta=6.0$
using varying values of $\kappa$.
In Figs.~\ref{Fig11-lattice-mpi-640}a and \ref{Fig11-lattice-mpi-640}b
and \ref{Fig11-lattice-mpi-640}c the form factors
$A_2^{u+d}(t)=A^Q(t)\equiv M_2^Q(t)$ and
$B_2^{u+d}(t)=B^Q(t)\equiv 2J^Q(t)-M_2^Q(t)$ and
$C_2^{u+d}(t)=C^Q(t)\equiv \frac15d_1(t)$ are shown for $m_\pi=640\,{\rm MeV}$
\cite{Gockeler:2003jf}. The results refer to a scale of $\mu=2\,{\rm GeV}$,
and cover the regions $-t\in[0,2.8]\,{\rm GeV}^2$ for $A^Q(t)$ and
$[0.6,2.8]\,{\rm GeV}^2$ for $B^Q(t)$ and $C^Q(t)$.
For comparison we plot the CQSM results at the corresponding scale and value of
$m_\pi$. Also here we observe a satisfactory agreement with the lattice data
within model accuracy and/or within the statistical accuracy of the lattice data.

Notice that the presence of explicit kinematical factors of ${\cal O}(\Delta)$
in the case of $B^Q(t)$, and of ${\cal O}(\Delta^2)$ in the case of $C^Q(t)$, see
Eq.~(\ref{Eq-app:ff-of-EMT-alternative}) in App.~\ref{App:Alternative-definition},
practically amplifies the statistical errors of the form factors --- as can
be seen in Figs.~\ref{Fig10-lattice-mpi-900} and \ref{Fig11-lattice-mpi-640}.
This is also the reason why no results for $C^Q(t)$ were presented in the
exploratory study of Ref.~\cite{Hagler:2003jd}.

The value of $2J^Q(t)\equiv (A^Q+B^Q)(t)$ at $t=0$ is of particular interest
since it gives the percentage of the total nucleon spin carried by quarks
\cite{Ji:1996nm}.
The lattice calculations yield
\be\label{Eq:JQ-lattice}
    2J^Q(0) = \cases{
    0.60 \pm 0.07  & at $\mu=1.76\,{\rm GeV}$ extrapolated to the physical point
             \cite{Mathur:1999uf}   \cr
    0.70 \pm 0.20  & at $\mu\sim 1.8\,{\rm GeV}$ for $m_\pi\sim 0.8\,{\rm GeV}$
             \cite{Gadiyak:2001fe}  \cr
    0.66 \pm 0.07  & at $\mu=2\,{\rm GeV}$ extrapolated to the physical point
             \cite{Gockeler:2003jf} \cr
    0.682\pm 0.018 & at $\mu=2\,{\rm GeV}$ for $m_\pi=900\,{\rm MeV}$
             \cite{Hagler:2003jd}.}
\ee
In the CQSM we have $2J^Q(0)=1$ for any $m_\pi$ at the low scale of the model,
since there are only quarks and antiquarks to carry the nucleon angular momentum,
and they must, of course, carry $100\,\%$ of it. Considering evolution
one obtains
\be\label{Eq:JQ-model}
    2J^Q(0) \approx 0.75 \;\;\;
    \mbox{from CQSM at $\mu\approx(1.7-2.0)\,{\rm GeV}$,}
\ee
and $2J^G(0) \approx 0.25$.
The result (\ref{Eq:JQ-model}) is --- within model accuracy --- in good agreement
with the lattice data in (\ref{Eq:JQ-lattice}).

For other aspects in the context of the nucleon spin structure discussed from
the CQSM point of view the reader is referred to \cite{Wakamatsu:2005vk}.

\newpage

\section{\boldmath  Chiral extrapolation of lattice data}
\label{Sec-7:lattice-lessons}

\begin{figure*}[b!]
\begin{tabular}{cc}
  \hspace{-0.5cm}
  \includegraphics[width=8cm]{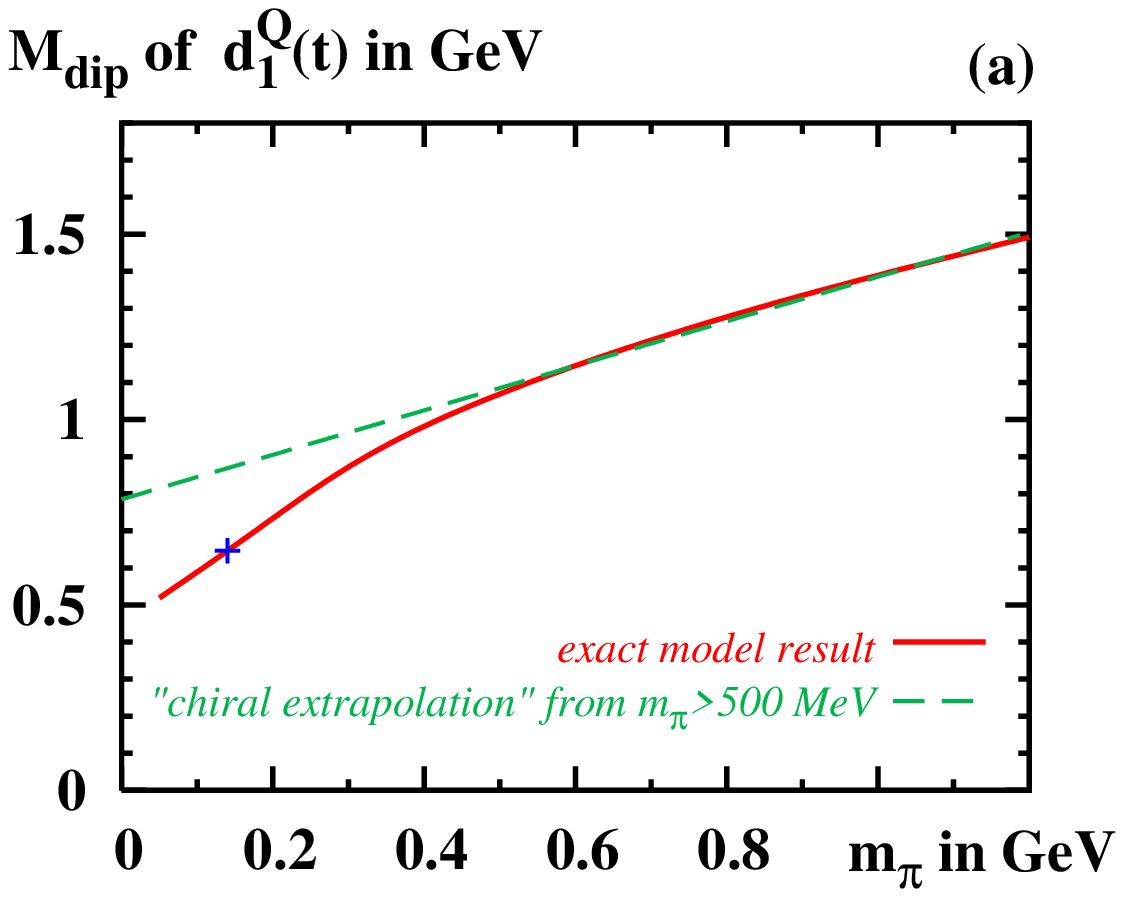} &
  \includegraphics[width=8cm]{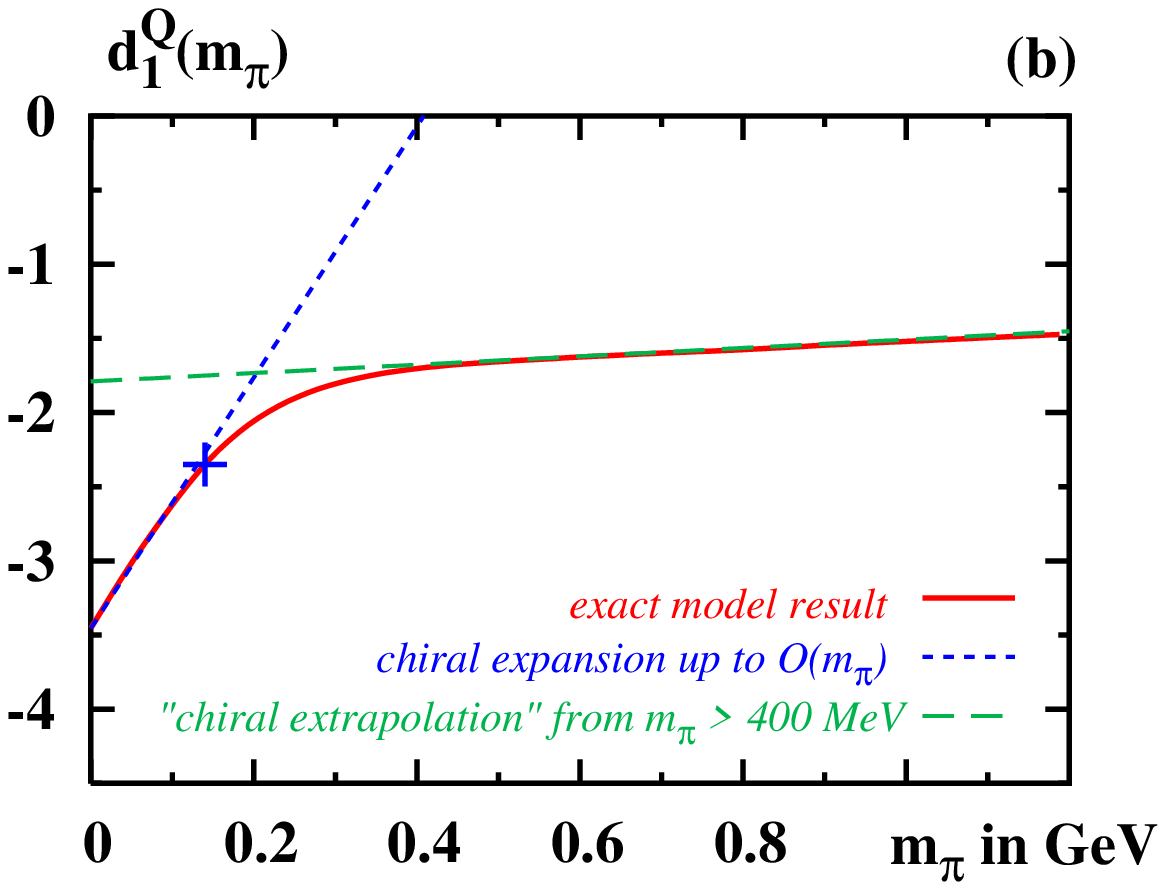}
\end{tabular}
    \caption{\label{Fig11-d1-dipol-mpi}
        \footnotesize\sl
    (a) The dipole mass $M_{\rm dip}$ of the form factor $d_1^Q(t)$ vs.\ $m_\pi$
    as obtained from the CQSM. The linear $m_\pi$-dependence in the region
    $m_\pi\gtrsim 500\,{\rm MeV}$ is emphasized.
    (b) The pion mass dependence of the form factor $d_1^Q(t)$ at zero
    momentum transfer as obtained from the CQSM at low scale.
    Solid line: The exact result from the CQSM.
    Dotted line: The chiral expansion of $d_1(m_\pi)$ to
    linear order in $m_\pi$ according to Eq.~(\ref{Eq:d1-vs-mpi}).
    Dashed line: The attempt of a ``chiral extrapolation''
    from the region of large $m_\pi \gtrsim 400\,{\rm MeV}$.
    In both figures the cross marks the physical point.}
\end{figure*}

Let us draw from our study some conclusions
which could be of interest in the context of the chiral extrapolation of lattice
data. Unfortunately, the model cannot provide any insights in this respect
concerning the $m_\pi$-dependence of $M_2^Q$ or $J^Q$. At the low scale
$M_2^Q=2J^Q=1$ which expresses consistently that the total nucleon momentum
and angular momentum in the model are carried by quarks --- irrespective
the value of $m_\pi$.
The model provides, however, interesting results for the $t$-dependence of
$M_2^Q(t)$, $J^Q(t)$ and $d_1^Q(t)$ including the normalization of the latter.

For that let us imagine that for some reason we were able to compute in the CQSM the
EMT form factors only for $m_\pi\gtrsim (400-500)\,{\rm MeV}$, and then forced to
extrapolate the results down to small $m_\pi$ in order to compare to the real
world --- using the model results from the large-$m_\pi$ region as main guideline.
What would we obtain?

First of all we emphasize that in the CQSM all form factors of the EMT can be
well approximated by dipoles. 
Though there is some prejudice in this respect based on the experience with
the electric and magnetic form factors of the proton, this is an important
and non-trivial observation.

Next, we consider the dipole masses of the form factors $M_2^Q(t)$ and $J^Q(t)$.
From the approximately linear $m_\pi$-behaviour of the respective dipole masses
in the $m_\pi\gtrsim (400-500)\,{\rm MeV}$ we might have been tempted to assume
also a linear behaviour in the region $m_\pi$-region down to the physical point.
Such an extrapolation would have resulted in a surprizingly accurate estimate
for the dipole masses of $M_2^Q(t)$ and $J^Q(t)$ at the physical point,
see Fig.~\ref{Fig8-d1-dipol-mpi}a.

However, this procedure would have failed in the case of the dipole mass of
$d_1^Q(t)$. In fact, also this dipole mass exhibits an approximately linear
$m_\pi$-behaviour in the region $m_\pi\gtrsim 500\,{\rm MeV}$,
see Fig.~\ref{Fig8-d1-dipol-mpi}a.
But presuming that this linear behaviour continues to hold also for
$m_\pi\lesssim 500\,{\rm MeV}$ would have resulted in a strong overestimate
of the dipole mass of $d_1(t)$ at the physical point of about $30\,\%$, see
Fig.~\ref{Fig11-d1-dipol-mpi}a.

Finally, let us discuss the $m_\pi$-dependence of the form factor $d_1^Q(t)$ at
zero momentum transfer. We observe an $m_\pi$-behaviour of $d_1^Q\equiv d_1^Q(0)$ in
the region of $m_\pi > 400\,{\rm MeV}$ which is linear to a very good approximation.
However, if we assumed that this linear $m_\pi$-dependence continued down to small
pion masses we would have underestimated the absolute value of $d_1^Q$
at the physical point by about $25\,\%$, and in the chiral limit by $50\,\%$,
see Fig.~\ref{Fig11-d1-dipol-mpi}b.

Recall that the chiral expansion of $d_1^Q$ to linear order in $m_\pi$ ---
i.e.\  up to the leading non-analytic contribution (in large $N_c$) according
to Eq.~(\ref{Eq:d1-vs-mpi}) --- approximates the full result rather well
up to the physical point. It is an interesting result that in our model $d_1^Q$
exhibits in the region of large $m_\pi > 400\,{\rm MeV}$ again a linear
$m_\pi$-dependence --- just as at low $m_\pi \lesssim 140\,{\rm MeV}$,
however, with a roughly 30 times smaller slope!

Of course, we make these observations in an effective chiral theory formulated in
the large $N_c$ limit. The situation could be different in full QCD simulations
performed at finite $N_c=3$. However, the model was observed  --- in spite of its
caveats like large vs.\ finite $N_c$ --- to provide a qualitatively good description
of lattice results on the pion mass dependence of the nucleon mass in the region
$300\,{\rm MeV}\lesssim m_\pi\lesssim 1.5\,{\rm GeV}$ \cite{Goeke:2005fs}.
Therefore it could be worth to keep in mind the lessons we learn here from
this model.

\newpage
\section{Conclusions}
\label{Sec-7:conclusions}

We have presented a study of the nucleon EMT form factors in the
large-$N_c$ limit in the framework of the CQSM in the region of
pion masses $140\,{\rm MeV}\le m_\pi\le 1.2\,{\rm GeV}$.
This work supplements the study of the nucleon EMT form factors in
the region $0 \le m_\pi\le 140\,{\rm MeV}$ \cite{accompanying-paper-I}.
There is a good reason why these studies have been presented separately.
The CQSM  \cite{Diakonov:yh,Diakonov:1987ty} has been used extensively
to study the nucleon at the physical point and in the chiral limit
with considerable phenomenological success
\cite{Christov:1995hr,Christov:1995vm,Silva:2005qm,Diakonov:1996sr,Diakonov:1997vc,Diakonov:1998ze,Wakamatsu:1998rx,Goeke:2000wv,Petrov:1998kf,Penttinen:1999th,Schweitzer:2002nm,Schweitzer:2003ms,Ossmann:2004bp,Wakamatsu:2005vk}.
The model owes its success to the facts that it is a QCD-inspired
field theoretic approach \cite{Diakonov:1983hh,Diakonov:1985eg,Diakonov:1995qy}
which correctly describes chiral symmetry playing an essential role in the
description of the nucleon.

Here we studied the CQSM at model parameters corresponding to large pion
masses far above the physical point, i.e.\  far away from the chiral limit.
Is the model applicable in this situation?
First exploratory studies in this direction have indicated a positive answer
\cite{Schweitzer:2003sb,Goeke:2005fs}. In order to shed further light to this
question we have focussed here on the nucleon form factors of the EMT
which is a central quantity for any field theoretic approach.

In our study we have demonstrated that the soliton solutions  for large $m_\pi$
obtained numerically in \cite{Goeke:2005fs} correspond to stable solitons, which
is an important prerequisite for the applicability of the approach.
We have shown that the model results for the energy density, the angular momentum
density, and the distributions of strong pressure and shear forces are in
qualitative agreement with what one expects as the constituents building
up the nucleon become heavy and the range of the pion cloud diminishes.
Namely, the spatial extension of the nucleon shrinks, the energy density increases,
the absolute values of the forces inside the nucleon increase
--- resulting in a smaller and more tightly bound nucleon.
These densities are related to the form factors of the EMT
\cite{Polyakov:2002yz}.

In order to test the model results in a more quantitative way we have compared
them to results for the EMT form factors from lattice QCD simulations
\cite{Mathur:1999uf,Gadiyak:2001fe,Hagler:2003jd,Gockeler:2003jf,Negele:2004iu}
performed at lattice parameters corresponding to $m_\pi\gtrsim 500\,{\rm MeV}$.
We observe a good agreement with the lattice data within an accuracy of $(10-30)\%$.
Also at the physical point the model was observed to work to within a similar accuracy
\cite{Christov:1995hr,Christov:1995vm,Silva:2005qm,Diakonov:1996sr,Diakonov:1997vc,Diakonov:1998ze,Wakamatsu:1998rx,Goeke:2000wv}.

The good performance of the model at the physical point and in the region
of large pion masses raises the question whether the model results could be
used as a guideline for the chiral extrapolation of lattice data.
Given the generally observed accuracy of the model it is clear
that it cannot be used as a precision extrapolation tool.
However, the model results can be helpful in two respects.

First, our results can be of use as qualitative guidelines for the
extrapolation of lattice data. For example, the model results indicate
that the EMT form factors can be well approximated by dipole fits, and that the
dipole masses of $M_2^Q(t)$ and $J^Q(t)$ exhibit to a good approximation a linear
$m_\pi$-dependence from $m_\pi = 1.2\,{\rm GeV}$ down to the physical point.
But our results indicate also that the chiral extrapolation of the form factor
$d_1^Q(t)$ from the region $m_\pi\gtrsim (400-500)\,{\rm MeV}$ could be a subtle
and difficult task.

Second, the success of the model at large pion masses could
provide a justification for applying the ``pion cloud'' idea
to the description of nucleon up to pion masses of ${\cal O}(1\,{\rm GeV})$
as explored in the finite range regulator approach. This approach may provide
a useful and precise tool for the chiral extrapolation of lattice data
as argued in \cite{Leinweber:1998ej,Thomas:2005rz}.

To conclude, we observe that the CQSM provides a consistent description
of the nucleon at large pion masses indicating that effective quark and
pion degrees of freedom can account for the gross features of the nucleon
properties also in this regime. Work which may provide further insigths
in this direction is in progress.

\vspace{0.3cm}
\noindent
{\bf Acknowledgements} \hspace{0.2cm}
We thank Philipp H\"agler, Pavel Pobylitsa, Maxim Polyakov and especially
Wolfram Schroers for fruitful discussions and valuable comments.
This research is part of the EU integrated infrastructure initiative
hadron physics project under contract number RII3-CT-2004-506078,
and partially supported by the Graduierten-Kolleg Bochum-Dortmund
and Verbundforschung of BMBF.
A.~S.\  acknowledges support from GRICES and DAAD.

\appendix

\section{\boldmath Alternative definition of form factors}
\label{App:Alternative-definition}

By means of the Gordon identity $2M_N\bar u^\prime\gamma^\alpha u=$
$\bar u^\prime(i\sigma^{\alpha\kappa}\Delta_\kappa+2P^\alpha)u$ one can rewrite
(\ref{Eq:ff-of-EMT}) as (see e.g.\ in Ref.~\cite{Ji:1996nm})
\ba
    \la p^\prime| \hat T_{\mu\nu}^{Q,G}(0) |p\rangle
    &=& \bar u(p^\prime)\biggl[
    A^{Q,G}(t)\,\frac{\gamma_\mu P_\nu+\gamma_\nu P_\mu}{2}+
    B^{Q,G}(t)\,\frac{i(P_{\mu}\sigma_{\nu\rho}+P_{\nu}\sigma_{\mu\rho})
    \Delta^\rho}{4M_N}
    \nonumber \\
    &+& C^{Q,G}(t)\,\frac{\Delta_\mu\Delta_\nu-g_{\mu\nu}\Delta^2}{M_N}
    \pm \bar c(t)g_{\mu\nu} \biggr]u(p)\, ,
    \label{Eq-app:ff-of-EMT-alternative} \ea
where
$A^{Q,G}(t)           = M_2^{Q,G}(t)$,
$A^{Q,G}(t)+B^{Q,G}(t)= 2\,J^{Q,G}(t)$,
$C^{Q,G}(t)           = \frac15\,d_1^{Q,G}(t)$.
In this notation the constraints (\ref{Eq:M2-J-d1}) read $A^Q(0)+A^G(0)=1$
and $B^Q(0)+B^G(0)=0$. The latter constraint is sometimes rephrased as the
"vanishing of the total nucleon gravitomagnetic moment".

\newpage


\end{document}